\documentclass[journal]{vgtc}         

\ifpdf
  \pdfoutput=1\relax                   
  \usepackage{graphicx}                
  \DeclareGraphicsExtensions{.pdf,.png,.jpg,.jpeg} 
\else
  \usepackage{graphicx}                
  \DeclareGraphicsExtensions{.eps}     
\fi%

\graphicspath{{figures/}{pictures/}{images/}{./}} 

\usepackage{amssymb}
\usepackage{microtype}                 
\PassOptionsToPackage{warn}{textcomp}  
\usepackage{textcomp}                  
\usepackage{mathptmx}                  
\usepackage{times}                     
\usepackage{cite}                      
\usepackage{tabu}                      
\usepackage{booktabs}                  
\usepackage{subfigure}
\usepackage[english, ruled, vlined, boxed, lined, linesnumbered]{algorithm2e}

\onlineid{109}

\vgtccategory{Research}
\vgtcpapertype{System}

\title{The Topology ToolKit}

\author{Julien Tierny,
        Guillaume Favelier,
        Joshua A. Levine, \textit{Member, IEEE},
        Charles Gueunet, and
        Michael Michaux}
\authorfooter{
\item
Julien Tierny, Guillaume Favelier, Charles Gueunet and Michael Michaux are 
with Sorbonne Universites, UPMC Univ Paris 06, CNRS, LIP6 UMR 7606, 
France.\protect~E-mails: \{julien.tierny, guillaume.favelier, 
charles.gueunet, michael.michaux\}@lip6.fr.
\item 
Joshua A. Levine is with the University of Arizona, USA.
\protect~E-mail: josh@email.arizona.edu.
\item 
Charles Gueunet is  also with Kitware SAS, France.
\protect~E-mail: charles.gueunet@kitware.com.}

\shortauthortitle{Tierny \MakeLowercase{\textit{et al.}}: 
the Topology ToolKit (TTK)}

\newcommand{\figref}[1]{Fig.~\ref{#1}}

\newcommand{\secref}[1]{Sec.~\ref{#1}}
\newcommand{\tabref}[1]{Tab.~\ref{#1}}
\newcommand{\algoref}[1]{Alg.~\ref{#1}}

\newcommand{\imageCaption}[1]{
\vspace{-4ex}
\caption{#1}\vspace{-1.5ex}}
\newcommand{\tableCaption}[1]{\caption{#1}}
\newcommand{\shrinkedSection}[1]{
\section{#1}
\vspace{-0.5ex}
}
\newcommand{\shrinkedSubSection}[1]{
\vspace{-0.5ex}
\subsection{#1}
\vspace{-0.5ex}}

\newcommand{\julien}[1]{#1}

\abstract{Topology-based techniques for data analysis and visualization have
recently grown in popularity, thanks to their ability to robustly and
efficiently extract features of interest at multiple scales of
importance.
Despite this success, this class of algorithms has not managed to make
it to end users without the dedicated support of an analyst.
We present a software platform for delivering topological analysis
techniques called the Topology ToolKit (TTK).
This system focuses on bridging two groups: (1) end users of topological
data analysis techniques and (2) developers interested in implementing
such techniques with minimal effort.  
For end users, TTK exposes topological data analysis at three different
entry points, each suitable for a different level of coding expertise.
For developers of new algorithms, we provide a modular framework of
convenience classes that help manage input data structures and reuse
rendering components.
Taking advantage of these features only requires the implementation of a
VTK wrapper class that is then mapped to a ParaView plugin. 
Our software system thus handles a broad set of use cases for multiple
stakeholders while still being practical and extensible.
TTK is necessarily open source, and we believe it is the first effort to
distribute topological data analysis to the masses.  
} 

\abstract{
This system paper presents the Topology ToolKit (TTK), a software platform
designed for 
topological data analysis in scientific visualization.
While topological data analysis has gained in popularity over the last two 
decades,
it has not yet been widely adopted as a standard data analysis tool for end 
users or developers.
TTK aims at addressing this problem by providing a unified, generic, 
efficient, and robust 
implementation of key algorithms 
for the topological analysis of scalar data,
including:
critical points, integral lines, persistence diagrams, persistence curves, 
merge trees, contour trees, Morse-Smale complexes, fiber surfaces, 
continuous 
scatterplots, 
Jacobi sets, Reeb spaces, and more. TTK is easily accessible to 
end users due to a tight integration with ParaView. 
It is also easily 
accessible to developers through a variety of bindings (Python, VTK/C++) for 
fast prototyping or through
direct, dependence-free, C++, to ease integration into 
pre-existing complex systems.
While developing TTK, we faced several algorithmic and software engineering 
challenges, which we document in this paper. In particular, we present an
algorithm for the construction of a discrete gradient that complies to the 
critical points extracted in the piecewise-linear setting. This algorithm 
guarantees a combinatorial consistency across the topological abstractions 
supported by TTK, and importantly, a unified implementation of topological 
data simplification for multi-scale exploration and analysis.
We also present a  cached triangulation data structure, 
that supports
time efficient and generic traversals, which self-adjusts its memory usage on 
demand for input simplicial
meshes and which implicitly emulates a triangulation for regular grids with 
no memory overhead. Finally, we describe an original software architecture, 
which guarantees memory efficient and direct accesses to TTK features, 
while still allowing for researchers powerful and easy bindings and extensions.
TTK is open source (BSD license) and its code, online documentation and video 
tutorials are available on TTK's website \cite{ttk}.
}

 
\CCScatlist{ 
 \CCScat{K.6.1}{Management of Computing and Information Systems}%
{Project and People Management}{Life Cycle};
 \CCScat{K.7.m}{The Computing Profession}{Miscellaneous}{Ethics}
}

\teaser{
  \includegraphics[width=\linewidth]{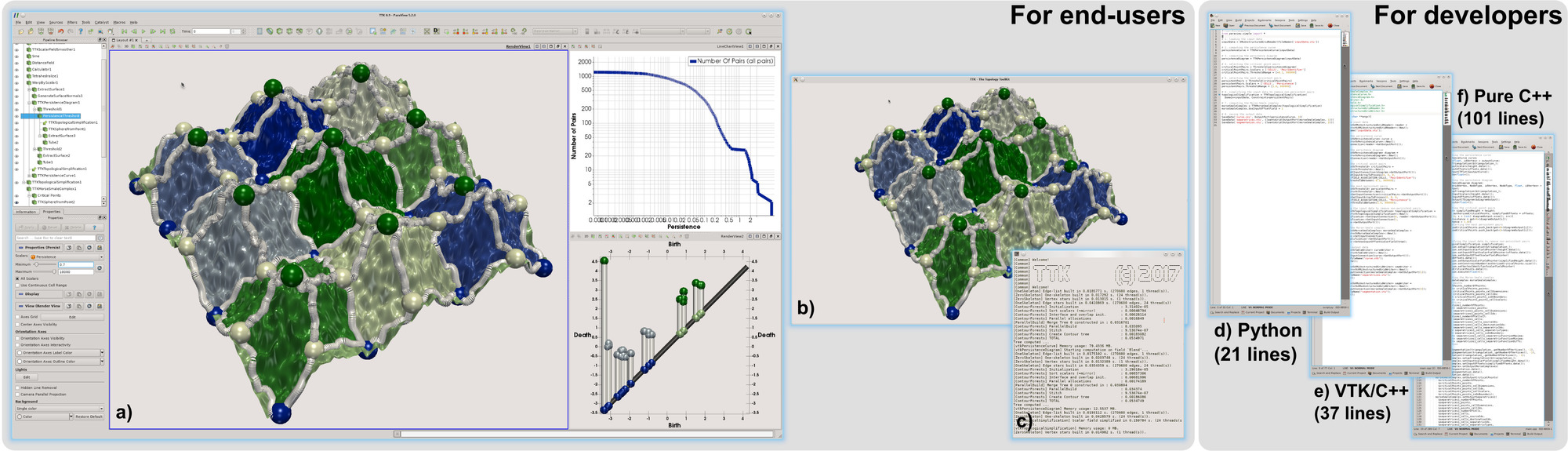}
  \vspace{-1ex}
  \vspace{-2.75ex}
  \caption{TTK is a software platform for 
topological data analysis in scientific visualization. It is both easily 
accessible 
to end users 
(collection of ParaView plugins (a), 
VTK-based generic GUIs (b) or command-line programs (c)) and flexible 
for developers
(Python (d), VTK/C++ 
(e) or 
dependence-free
C++ (f) bindings).
TTK provides an efficient and unified approach to topological data 
representation and 
simplification, 
which enables in this example a discrete Morse-Smale complex (a) to comply 
to the level of simplification dictated by a 
piecewise linear
persistence diagram
(bottom-right linked view, a).
Code snippets are provided (d-f) to reproduce this pipeline.
}
\vspace{-0.5ex}
	\label{fig_teaser}
}

\vgtcinsertpkg

\begin{document}


\firstsection{Introduction}

\maketitle

\label{sec_intro}
As scientific datasets become more intricate and larger in size, advanced 
data 
analysis algorithms are needed for their efficient visualization and 
exploration. For scalar field visualization, topological analysis techniques 
\cite{heine16}
have shown to be practical solutions in various contexts by enabling the 
concise and complete capture of the structure of the input data into high-level 
\emph{topological abstractions} such as contour trees \cite{carr00}, 
Reeb graphs \cite{pascucci07, biasotti08, tierny_vis09}, or Morse-Smale 
complexes \cite{gyulassy_vis08, Defl15}.
Such topological abstractions are fundamental data structures that 
enable 
the development of advanced data 
analysis, exploration and visualization techniques, including for instance: 
small seed set extraction for fast isosurface traversal \cite{vanKreveld97, 
carr04}, feature tracking \cite{sohn06}, transfer function design for volume 
rendering \cite{weber07}, similarity estimation \cite{thomas14}, or 
application-driven segmentation and analysis tasks \cite{laney_vis06, 
gyulassy07, gyulassy_ev14, gyulassy_vis15}. Successful applications in a 
variety of fields of science,
including combustion 
\cite{laney_vis06, bremer_tvcg11, gyulassy_ev14}, 
material sciences \cite{gyulassy07, gyulassy_vis15},
chemistry 
\cite{chemistry_vis14}, 
or astrophysics \cite{sousbie11, shivashankar2016felix} to name a few, have 
even been documented, which further stresses the importance of this class of 
techniques.
Despite this popularity and success in applications, topological data analysis 
(TDA) has not yet been widely adopted as a standard data analysis tool for end 
users and developers.
While some open source implementations for specific algorithms are available
\cite{libTourte, vijayRG, doraiswamyRecon, 
vtkReebGraph, vijayMSC, gerberMSR, 
sousbie11, Chen:ECG:2007}, we still identify three main issues preventing a 
wider adoption of TDA.

First, 
these implementations
lack, in general, support for standard data file 
formats, genericity regarding the dimensionality of the input data, 
integration into graphical user front ends, or access through high-level 
scripting languages. These limitations challenge their adoption by end users 
and domain experts with little or no programming knowledge.

Second, regarding software developers, each implementation comes with its own 
internal 
data structures or 
its own list of third-party software dependencies, 
which challenges their integration into pre-existing, complex systems
for 
visualization or data analysis.

Third, regarding researchers,
despite the isolated open source implementations mentioned above, 
many TDA algorithms do not have publicly available implementations,
which challenges reproducibility.
%
While other research communities have excelled at providing software platforms 
that ease the implementation, benchmarking, and distribution of research code 
(such as the \emph{Image Processing On Line} platform \cite{ipol}), to the best 
of our knowledge, there has not been such a federating initiative 
for TDA
codes in scientific visualization.

This system paper presents the Topology ToolKit (TTK) \cite{ttk}, a software 
platform 
for 
\julien{the topological analysis of scalar data}
in scientific visualization, which
\julien{addresses}
the three core problems described above:
\emph{(i)}
accessibility to 
end users,
\emph{(ii)}
flexibility for developers and 
\emph{(iii)}
ease of extension and 
distribution of new algorithms for researchers. 
TTK provides a unified, generic, 
efficient, and robust implementation of key algorithms for the 
topological analysis of scalar data.
It is easily accessible to end users thanks to a tight 
integration with ParaView (\figref{fig_teaser}, left) and
flexible for developers (\figref{fig_teaser}, right) through a 
variety of bindings (Python, VTK/C++) or direct, third-party 
\julien{dependency}-free, 
C++ access (to ease integration in pre-existing complex systems). Finally, it 
facilitates the implementation, integration, and distribution of TDA codes,
by simply requiring the implementation of a handful of 
functions, while providing efficient data structures and, thanks to ParaView, 
advanced IO, rendering and interaction support for end users. 

While developing TTK, we faced several algorithmic and software engineering 
challenges, which we document in this paper.

\noindent
  \textbf{(i) Algorithmic consistency:} For advanced analysis tasks, it 
can be desirable to combine several topological abstractions 
\cite{laney_vis06,chemistry_vis14}. However, each
abstraction 
comes with its 
own 
simplification mechanism, which challenges the development of a unified
framework. More important, several competing formalisms exist to 
represent the input data, namely the piecewise-linear setting 
\cite{banchoff70, edelsbrunner09} and the Discrete Morse Theory setting 
\cite{forman98}. The lack of compatibility between these two representations 
challenges 
even more the design of a unified framework.

\noindent
  \textbf{(ii) Core data structures:} Combinatorial algorithms for TDA
mostly involve mesh traversal routines. Thus, 
generic and time efficient triangulation data structures must be derived.
  
  \noindent
  \textbf{(iii) Software engineering:} Designing a software library which has 
no 
third-party dependency and which also seamlessly integrates into a 
complex visualization system such as ParaView is challenging. Related 
challenges include avoiding data copy 
within
the visualization 
pipeline. 
Also, designing such a flexible library in a way that still 
enables easy extensions is an additional difficulty. 

\newcommand{\domain}{\mathcal{M}}
\newcommand{\range}{\mathbb{R}}
\newcommand{\st}[1]{St(#1)}
\newcommand{\simplex}{\sigma}
\newcommand{\lk}[1]{Lk(#1)}
\newcommand{\lkminus}[1]{Lk^-(#1)}
\newcommand{\lkplus}[1]{Lk^+(#1)}
\newcommand{\sub}[1]{f^{-1}_{-\infty}(#1)}
\newcommand{\sur}[1]{f^{-1}_{+\infty}(#1)}
\newcommand{\Path}{P}
\newcommand{\offset}{\mathcal{O}}
\newcommand{\persistenceDiagram}{\mathcal{D}}
\newcommand{\Index}{\mathcal{I}}
\newcommand{\persistenceCurve}{\mathcal{C}}
\newcommand{\reebGraph}{\mathcal{R}}
\newcommand{\morseSmale}{\mathcal{MS}}
\newcommand{\pointList}{\mathcal{L}_P}
\newcommand{\simplexList}{\mathcal{L}_S}

\subsection*{Contributions}
This paper makes the following new contributions:
\begin{enumerate}
  \item{
  \vspace{-2ex}
  \emph{An algorithm} (\secref{sec_topologicalSimplification})
  to construct a discrete gradient which complies to the critical points 
extracted  in the piecewise linear (PL) setting.
Each critical simplex resulting from this algorithm is 
located in the star of a PL critical point. This 
relationship between the discrete and PL settings enables 
a 
combinatorial consistency across the different topological abstractions 
supported by TTK. As a byproduct, it allows for a unified and independent 
topological 
simplification procedure for multiscale exploration and analysis.}
  \item{
  \vspace{-2ex}
  \emph{A data structure} (\secref{sec_triangulation}) for time efficient 
traversals on 2D or 
3D piecewise linear triangulations. In the case of input meshes, it 
self-adjusts its memory footprint depending on the traversal operations 
it is queried for. In the case of regular grids, it implicitly emulates 
a triangulation with no memory overhead.}
  \item{
  \vspace{-2ex}
  \emph{A software architecture} (\secref{sec_software})
  that eases the development and distribution of 
  TDA
code to end users.
The creation of a new module only requires the implementation of a handful of 
functions, while TTK automatically generates a command-line program, a 
VTK-based GUI
and a ParaView plugin connected to the module. 
}
  \item{
  \vspace{-2ex}
  \emph{A software collection} (\secref{sec_collection})
  that implements
  in 
a unified and generic  way a variety of
TDA
algorithms. It is accessible to end users as 
command line programs, VTK-based GUIs, or ParaView 
plugins. It is accessible to developers through a variety of bindings: Python, 
VTK/C++, or \julien{dependency}-free C++. }
\end{enumerate}

\shrinkedSection{Related work}
\label{sec_relatedWork}
In this section, we discuss three main categories of prior work related to 
this paper: existing visualization front ends, TDA software packages, 
and triangulation data structures for TDA. 

\shrinkedSubSection{Visualization front ends}
In this subsection,
we briefly mention existing efforts to enhance the accessibility of
visualization to end users.  Many of the successes in this area are
either specific libraries and toolkits, such as the Visualization
ToolKit (VTK)~\cite{vtkbook} and the Insight ToolKit
(ITK)~\cite{itkbook}.  TTK is similar to these libraries albeit with a 
specific focus on topological data analysis.
Related, largely turnkey tools 
often deliver features
of such toolkits with richer interfaces.  
Examples 
include
ParaView~\cite{paraview}, VisIt~\cite{HPV:VisIt},
VisTrails~\cite{BavoilCSVCSF05}, SCIRun~\cite{SCI:SCIRun},
MeVisLab~\cite{mevislab}, and Amira~\cite{amira}. 
Many of the above systems employ a dataflow network as a key component
to their visualization pipeline~\cite{moreland2013survey}.  
Extensions of this model have been proposed for specific use-cases or 
application areas, for instance
for higher order finite elements~\cite{schroeder2006methods}.  
TTK also 
differs
sharply 
from alternate forms of specializations, such as 
domain-specific
languages for visualization that have been recently developed, including
\emph{Diderot}~\cite{kindlmann2016diderot}, 
\emph{Scout}~\cite{mccormick2007scout},
\emph{Vivaldi}~\cite{choi2014vivaldi}, and 
\emph{ViSlang}~\cite{rautek2014vislang}.

\shrinkedSubSection{TDA software packages}
Existing TDA software packages can be
categorized into two groups.
While their
inputs
and outputs 
greatly differ, they share the common
goal of extracting features which capture topological invariants.


\begin{figure*}
  \vspace{-1ex}
  \includegraphics[width=\linewidth]{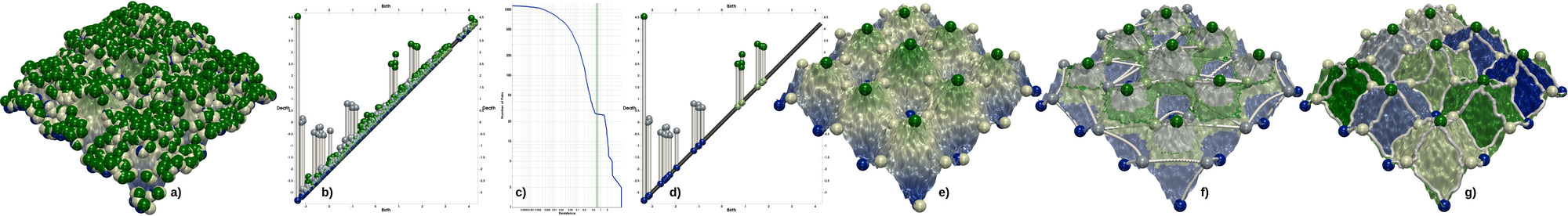}
  \vspace{-0.5ex}
  \imageCaption{
  Height function $f$ (blue to green) on a noisy terrain. (a) 
Initially, $f$ admits many critical points (spheres, blue: minima, white: 
saddles, green: maxima), resulting in a persistence diagram 
$\persistenceDiagram(f)$  with many small bars near the diagonal (b). The 
persistence curve $\persistenceCurve(f)$  exhibits a clear plateau (c) 
separating noise from features (green vertical line). Our approach to 
topological simplification pre-simplifies $f$ to the corresponding persistence 
threshold in a unified way, which is consistently interpreted by 
all the supported topological abstractions: persistence diagrams (d), critical 
points (e), contour trees (f) and even discrete Morse-Smale complexes (g). Note
that any application-driven metric could be used in place of persistence.}
\vspace{-0.5ex}
  \label{fig_simplification}
\end{figure*}

\noindent
\textbf{TDA on low-dimensional manifolds:}
\label{sec:fieldrefs}
The first category 
focuses on 
topological 
abstractions of scalar data on low-dimensional manifolds (typically in 2D or 
3D) for the purpose of data analysis and visualization as discussed in Sec.~1. 
TTK primarily targets these applications.
Such topological abstractions include for instance
critical points \cite{banchoff70},
persistence diagrams \cite{edelsbrunner02, cohen-steiner05},
Reeb graphs \cite{reeb46, 
pascucci07, biasotti08, tierny_vis09, doraiswamy12,parsa12, doraiswamy13} (and 
their loop-free variant, contour trees \cite{boyell63, tarasov98, carr00}), 
and Morse-Smale complexes \cite{edelsbrunner03b, gyulassy_vis08, robins11, 
ShivashankarN12, Defl15}. 
Dillard implements \emph{libtourte}, a library
computing the contour 
tree~\cite{libTourte}, while Doraiswamy et al.'s \emph{libRG}
library~\cite{vijayRG} and \emph{Recon}~\cite{doraiswamyRecon} as
well as Tierny's \emph{vtkReebGraph}~\cite{vtkReebGraph} compute the Reeb
graph.  
Shivashankar 
and Natarajan have focused on a scriptable implementation
of
the Morse-Smale complex~\cite{vijayMSC} based on
their parallel algorithms 
in 2D~\cite{ShivashankarMN12} and 3D~\cite{ShivashankarN12}.  
Sousbie developed \emph{DisPerSE}, an
implementation of the Morse-Smale complex focused on cosmology data 
analysis~\cite{sousbie11}.
Finally, Chen et al. provide an
implementation of the Morse Decomposition for 2/2.5D vector
fields~\cite{Chen:ECG:2007}.

Although powerful, the above tools 
often 
lack the level of integration 
required for end users' adoption.
Most of them come with custom file formats 
\cite{libTourte, vijayRG, doraiswamyRecon, sousbie11, Chen:ECG:2007}. This 
forces users to write data converters for each tool, which 
greatly impairs
adoption by end users, who can be domain experts with no 
programming background. Also, these tools often  come only as 
libraries or command-line programs, 
which can also discourage end users.
In contrast, some tools \cite{vtkReebGraph, 
gerberMSR} directly rely on established toolkits such as
the Visualization ToolKit 
(VTK) \cite{vtkbook} or the R statistical environment \cite{r} and thus 
benefit 
from a rich support for most standard data file formats.
However, 
the dependency on these complex environments has not been designed in these 
tools to be optional. This constitutes a serious limiting factor for developers 
of pre-existing complex systems, who often want to minimize the number of 
third-party dependencies their system rely on. In contrast, TTK's software 
architecture (\secref{sec_software}) has been specifically designed such 
that TTK can be called by \julien{dependency}-free C++ codes,
by using primitive types 
only.
This allows  integrating TTK 
seamlessly in any system written in C++, without having to pull any extra 
third-party dependency.
Optionally, TTK can wrap around both VTK 
and 
ParaView 
to leverage their rich IO support as well as their 
powerful user interface capabilities. Finally, existing tools often lack 
genericity in terms of input data representation or dimensionality. For 
instance, the two Morse-Smale complex implementations by 
 Shivashankar 
and Natarajan \cite{vijayMSC} 
are either designed for 
triangulations in 2D 
or regular grids in 3D. In contrast, TTK supports in 
a unified way both representations in both dimensions. 
Also, TTK is based on a tighter integration with ParaView, which allows 
end users without any programming skill 
to easily interact with it, without even having to use scripting languages such 
as Python.




\noindent
\textbf{TDA on high-dimensional point clouds:}
A second set of TDA software packages rather focus on estimating persistent 
homology \cite{edelsbrunner02,zomorodian2005computing, edelsbrunner09} on point 
clouds, usually for topology inference applications in high dimensions. 
One of the earliest implementations 
is
\emph{Mapper}~\cite{singh2007topological}, 
which is used
by the commercial product
\emph{Ayasdi}~\cite{ayasdiURL}.  
Both \emph{Dionysus}~\cite{Morozov:Dionysus} 
and \emph{JavaPlex}~\cite{AdamsTV14} 
provide implementations of the
standard algorithm by Zomorodian and Carlsson 
\cite{zomorodian2005computing}.  \emph{Dionysus} also implements
persistent cohomology~\cite{de2011persistent} and zigzag persistent
cohomology~\cite{carlsson2009zigzag}.  \emph{Perseus}~\cite{Nanda:Perseus}
implements a preprocessing procedure that reduces the number of filtered
cells in the standard algorithm~\cite{MischaikowN13}.

More recent implementations focus on variations, either in the domain or
outputs. 
Gerber et al.  implement
\emph{MSR}~\cite{gerberMSR} for approximating Morse-Smale complexes on
k-nearest neighbor graphs of high-dimensional
data~\cite{gerber2010visual}.
\emph{Gudhi}~\cite{MariaBGY14} 
implements persistent homology on
simplicial complexes relying on the \emph{Simplex tree}~\cite{BoissonnatM14}
data structure.  \emph{Phat}~\cite{BauerKRW14} 
relies on
matrix reduction operators for efficient computations.  Fasy et al.'s TDA 
package provides an interface to
\emph{Gudhi}, \emph{Dionysus}, and \emph{Phat} in the popular analysis language
R~\cite{r,FasyKLM14}.  \emph{SimPers} computes persistent homology on simplicial
maps~\cite{dey2014computing}.  Bubenik's persistence landscapes
toolbox~\cite{bubenik2017persistence} computes a more descriptive statistical 
summary of homology than the typical persistence diagram.
In contrast to these packages, TTK specifically targets low dimensional (2D 
or 3D) domains for applications in scientific data analysis and visualization. 



\shrinkedSubSection{Triangulation data structures for TDA}
\label{sec_relatedTriangulation}
Combinatorial TDA algorithms mostly involve mesh traversal routines. Therefore, 
corresponding implementations must rely on  data structures 
providing time efficient traversal queries.
Data structures for triangulations is a well researched 
topic \cite{FlorianiH05,joy2003data}
and a variety of approaches exist for storing them with various 
sorts of trade-offs between memory footprint and time efficiency.  
Some data structures, such as \emph{OpenMesh} \cite{BSBK02}, 
\emph{surface\_mesh} \cite{SiegerB11} or \emph{SimplexMesh} 
\cite{Batty:SimplexMesh}, employ variants of
the half-edge  structure~\cite{Weiler85}.
In contrast,  several other
mesh libraries employ a
cell-based representation, where only points and cells of highest dimension 
are stored.
For instance, VTK \cite{vtkbook} implements this strategy with its mesh 
data structures,
whose design clearly trades efficiency for generality as these structures 
support 
arbitrary polyhedra and polygons.
CGAL's~\cite{cgal:eb-16b} 2D and
3D triangulation data structures~\cite{cgal:py-tds2-16b,
cgal:pt-tds3-16b}, Mesquite~\cite{BrewerDKLM03}, and
VCGLib~\cite{Cignoni:VCGLib}, the underlying library behind
MeshLab~\cite{CCCDGR08}, all employ cell-based data structures.  

A major drawback of these cell-based implementations is that the connectivity 
of cells of intermediate dimensions (edges and triangles in tet-meshes) 
must be re-computed upon each query. However, such queries 
are common practice in TDA (\secref{sec_specification}). This drawback is  
attenuated by data structures which aim at balancing time and memory efficiency 
\cite{BoissonnatM14,BoissonnatST15,WeissFFV11,WeissIFF13,Zomorodian10} but it 
is accentuated by data structures which further compress the adjacency 
information \cite{BlandfordBCK05,GurungLLR13, GurungLLR11, 
LuffelGLR14}.
In contrast, TTK 
implements a cached triangulation data structure 
(\secref{sec_triangulation}), which provides lookup-based time-efficient 
queries,
while self-adjusting its memory footprint on demand, 
and implicitly emulating 
triangulations in the case of regular grids.

\shrinkedSection{Preliminaries}
\label{sec_preliminaries}
This section briefly describes our formal setting. We refer the reader to 
reference books for introductions to Morse theory \cite{milnor63}, 
computational topology \cite{edelsbrunner09} and Discrete Morse Theory 
\cite{forman98}.


\shrinkedSubSection{Input data}
\label{sec_preliminariesInput}
Without loss of generality, we assume that the input data is a piecewise 
linear (PL) scalar field $f : \domain \rightarrow \mathbb{R}$ defined on a 
PL $d$-manifold $\domain$ with $d$ equals 2 or 3. It has 
value at the set of vertices $\domain^0$ of $\domain$ and is linearly 
interpolated on the simplices 
of higher dimension. 
Adjacency relations on $\domain$ can be described in a dimension independent 
way. 
The \emph{star} $\st{\simplex}$ of a simplex $\simplex$ is the set of 
simplices 
of $\domain$ which contain $\simplex$ as a face. The \emph{link} 
$\lk{\simplex}$ is 
the set of faces of the simplices of $\st{\simplex}$ which do not intersect 
$\simplex$. 
In the following,
the topology of $\domain$
will be mostly described in terms of its 
\emph{Betti numbers} $\beta_i$ (the ranks of its homology groups 
\cite{edelsbrunner09}), which 
correspond in 3D to 
the numbers of connected 
components ($\beta_0$), non collapsible cycles ($\beta_1$) and  voids 
($\beta_2$).

\shrinkedSubSection{Geometric features}
For visualization and data analysis purposes, several low-level geometric
features can be defined given the input data. Given an isovalue $i \in \range$, 
the \emph{level set}, noted $f^{-1}(i)$, is the pre-image of $i$ onto $\domain$ 
through $f$: $f^{-1}(i) = \{ p \in \mathcal{M} ~ | ~ f(p) = i \}$. The 
\emph{sub-level set}, noted $\sub{i}$, is defined as the pre-image of the open 
interval $(-\infty, i)$ onto $\domain$ through $f$: 
$\sub{i} = \{ p \in \mathcal{M} ~ | ~ f(p) < i \}$. Symmetrically, the 
\emph{sur-level set} $\sur{i}$ is defined by $\sur{i} = \{ p \in \mathcal{M} ~ 
| ~ 
f(p) > i \}$. 
An \emph{integral line} is a path on $\domain$ which is everywhere tangential 
to $\nabla f$. 
Topological data 
analysis can be seen as the study of the topological transitions
(in terms of Betti numbers) of these objects as one 
sweeps the range
$\range$ \cite{milnor63}.

\shrinkedSubSection{Critical points}
\label{sec_criticalPoints}
The points of $\domain$ where the Betti numbers of $\sub{i}$ change 
are the \emph{critical points} of $f$ 
(\figref{fig_simplification}(e)). Let $\lkminus{v}$ be 
the 
\emph{lower link} of the vertex $v$: $\lkminus{v} = \{ \sigma \in \lk{v} ~ | ~ 
\forall u \in \sigma : f(u) < f(v)\}$. The \emph{upper link} $\lkplus{v}$ is 
given 
by $\lkplus{v} = \{ \sigma \in \lk{v} ~ | ~ 
\forall u \in \sigma : f(u) > f(v)\}$. To classify $\lk{v}$
without ambiguity into either lower or upper links, the restriction of $f$ to 
the vertices of $\domain$ is assumed to be injective. This is easily enforced 
in practice by a variant of simulation of simplicity \cite{edelsbrunner90}. 
This is achieved by considering an associated injective integer offset 
$\offset(v)$, which initially typically corresponds to the vertex position 
offset in memory. Then, when comparing two vertices, if these share the same 
value $f$, their order is disambiguated by their offset $\offset$. A vertex $v$ 
is regular, if and only if both $\lkminus{v}$ and 
$\lkplus{v}$ are simply connected. Otherwise, $v$ is a \emph{critical 
point} of $f$ \cite{banchoff70}. Let $d$ be the dimension of $\domain$. 
Critical points can be classified with their \emph{index} $\Index$, which 
equals 0 for 
minima ($\lkminus{v} = \emptyset$), 1 for 1-saddles ($\beta_0(\lkminus{v}) = 
2$), $(d - 1)$ for $(d-1)$-saddles ($\beta_0(\lkplus{v}) = 2$) and $d$ for 
maxima ($\lkplus{v} = 
\emptyset$). Vertices for which $\beta_0(\lkminus{v})$ or
$\beta_0(\lkplus{v})$ are greater than 2 are called \emph{degenerate saddles}.
For bivariate functions $f: \domain \rightarrow \mathbb{R}^2$, 
the notion of \emph{Jacobi set} \cite{edelsbrunner04}  extends that of 
critical points \cite{tierny_vis16}. Bivariate analogs of level-sets 
(\emph{fibers}) also change their topology in their vicinity. 

\shrinkedSubSection{Topological persistence}
\label{sec_persistence}
The distribution of critical points of $f$ can be
represented by a first topological abstraction, called the \emph{persistence 
diagram}
\cite{edelsbrunner02, cohen-steiner05} 
(\figref{fig_simplification}(d))\julien{, which 
 also provides a measure 
 of topological noise on critical point pairs.} 
By applying the Elder's 
rule \cite{edelsbrunner09}, critical points can be arranged in a set of pairs, 
such that each critical point appears in only one pair $(c_i, c_j)$ with 
$f(c_i) < f(c_j)$ and $\Index(c_i) = \Index(c_j) - 1$. The persistence diagram 
$\persistenceDiagram(f)$ embeds each pair $(c_i, c_j)$ in the plane such 
that its horizontal coordinate equals $f(c_i)$, and the vertical coordinate of 
$c_i$ and $c_j$ are $f(c_i)$ and $f(c_j)$. The height of the pair $p = f(c_j) 
- f(c_i)$ is called the \emph{persistence}  and denotes the 
life-span of the topological feature created in $c_i$ and destroyed in $c_j$. 
In 
low dimensions, the persistence of the pairs linking critical points of index 
$(0,1)$, 
$((d-1),d)$ and $(1,2)$ (in 3D) denotes the life-span of  
connected components, voids and non-collapsible 
cycles of $\sub{i}$.
The \emph{persistence curve} $\persistenceCurve(f)$ plots the number of 
critical 
pairs as a function of their persistence
(\figref{fig_simplification}(c)).

\shrinkedSubSection{Reeb graphs}
\label{sec_preliminariesRG}
The \emph{Reeb graph} \cite{reeb46} 
segments $\domain$ into regions where the connectivity of $f^{-1}(i)$ does not 
change. 
Let $f^{-1}(f(p))_p$ be the connected component of $f^{-1}(f(p))$ 
containing $p$. The Reeb graph $\reebGraph(f)$ is a one-dimensional simplicial 
complex defined as the quotient space $\reebGraph(f) = \domain / \sim$ by the 
equivalence relation $p_1 \sim p_2$, which holds if $p_2 \in 
f^{-1}(f(p_1))_{p_1}$. 
For bivariate data $f : \domain 
\rightarrow \mathbb{R}^2$, Reeb graphs extend to
\emph{Reeb spaces} \cite{Edelsbrunner08, tierny_vis16}, being this time  
2D cell complexes.

Variants of the Reeb graph can be defined relative to the connected 
components of 
$\sub{i}$ and $\sur{i}$,
yielding the notion of \emph{merge tree} (specifically 
\emph{join} and \emph{split} trees 
for
 $\sub{i}$ and $\sur{i}$). In 2D, the persistence pairs of 
$\persistenceDiagram(f)$ can be efficiently computed from the 
join (split) tree, by pairing each saddle, in increasing (decreasing) order 
of $f$ values, with the highest (lowest) non-paired minimum (maximum) its 
contains in its sub-tree \cite{edelsbrunner09}.
The contour tree
(the loop-free 
variant of $\reebGraph(f)$
\julien{for simply connected domains})
can be 
efficiently computed 
by combining the join and split 
trees \cite{tarasov98, carr00} (\figref{fig_simplification}(f)).

\shrinkedSubSection{Morse-Smale complexes and Discrete Morse Theory} 
\label{sec_dmt}
The \emph{Morse-Smale complex} 
\cite{Defl15}
segments $\domain$ into regions where integral lines share both 
their origin
and 
destination
(\figref{fig_simplification}(g)).
Given a critical point $p$, its \emph{ascending} (resp.
\emph{descending}) \emph{manifold} is defined as the set of points belonging to 
integral lines whose origin (resp. destination) is $p$. 
The \emph{Morse complex} is 
the complex formed by all descending 
manifolds.
$f$ is said to be a \emph{Morse-Smale function} if it admits no 
degenerate saddle and if its ascending and descending manifolds only intersect 
transversally (if the codimension of their intersection 
equals the sum of their codimensions \cite{guillemin74}). Then, the 
\emph{Morse-Smale complex} $\morseSmale(f)$ is 
the complex formed by 
the intersection of the Morse complex of $f$ and that of $-f$. Concretely, all 
the points of a given cell (of arbitrary dimension) of  $\morseSmale(f)$ belong 
to integral lines having the same origin and destination.
In 3D, the persistence pairs 
corresponding to non-collapsible cycles 
can be efficiently extracted
by visiting the 2-saddles 
not already present in $\persistenceDiagram(f)$ (\secref{sec_preliminariesRG}) 
in ascending 
order and pairing each one  with the highest, non-paired, 1-saddle it is 
connected to through a 1-dimensional cell of $\morseSmale(f)$ (called 
\emph{saddle-connector}) 
and reverting the gradient along that cell.
The robust computation of Morse-Smale complexes for PL scalar fields has been a 
long time challenge for the community, as existing algorithms 
\cite{edelsbrunner03b} were highly complex and required many special cases to 
account for the transversal intersection condition in 3D.
Fortunately, 
an alternate formalism, namely \emph{Discrete Morse Theory} \cite{forman98} 
(DMT), enabled the definition of elegant and robust algorithms 
\cite{gyulassy_vis08, robins11, ShivashankarN12}, improving the applicability 
of Morse-Smale complexes.
We briefly describe here DMT 
in the case of PL manifolds, but it remains valid for arbitrary CW 
complexes.

\begin{figure}
  \centering
  \includegraphics[width=0.32\linewidth]{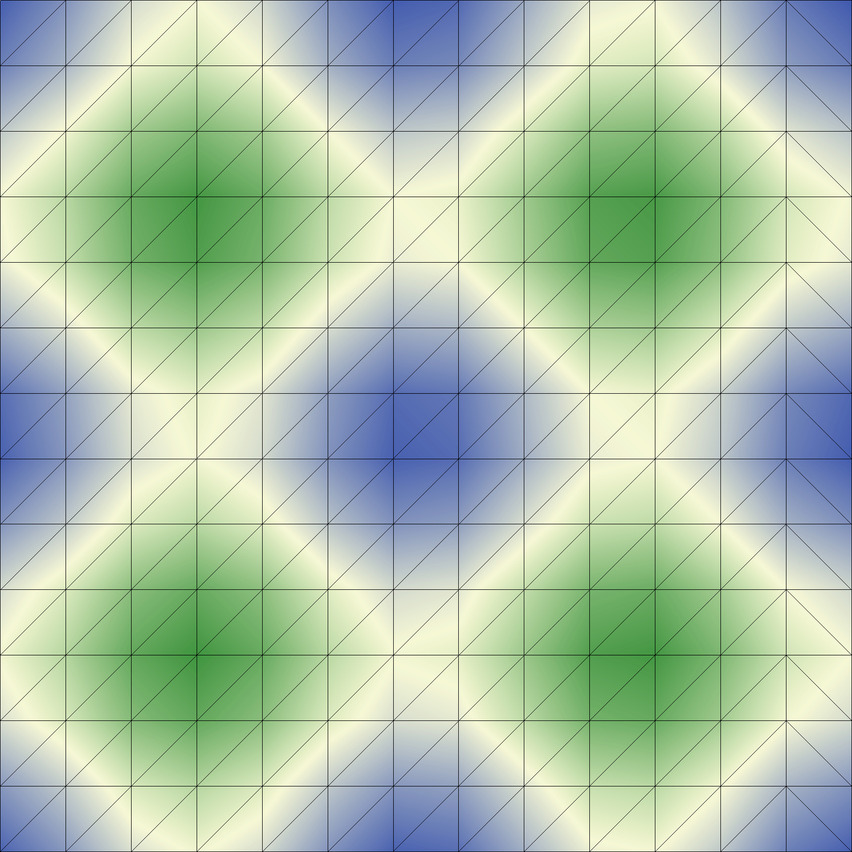}
  \hfill
  \includegraphics[width=0.32\linewidth]{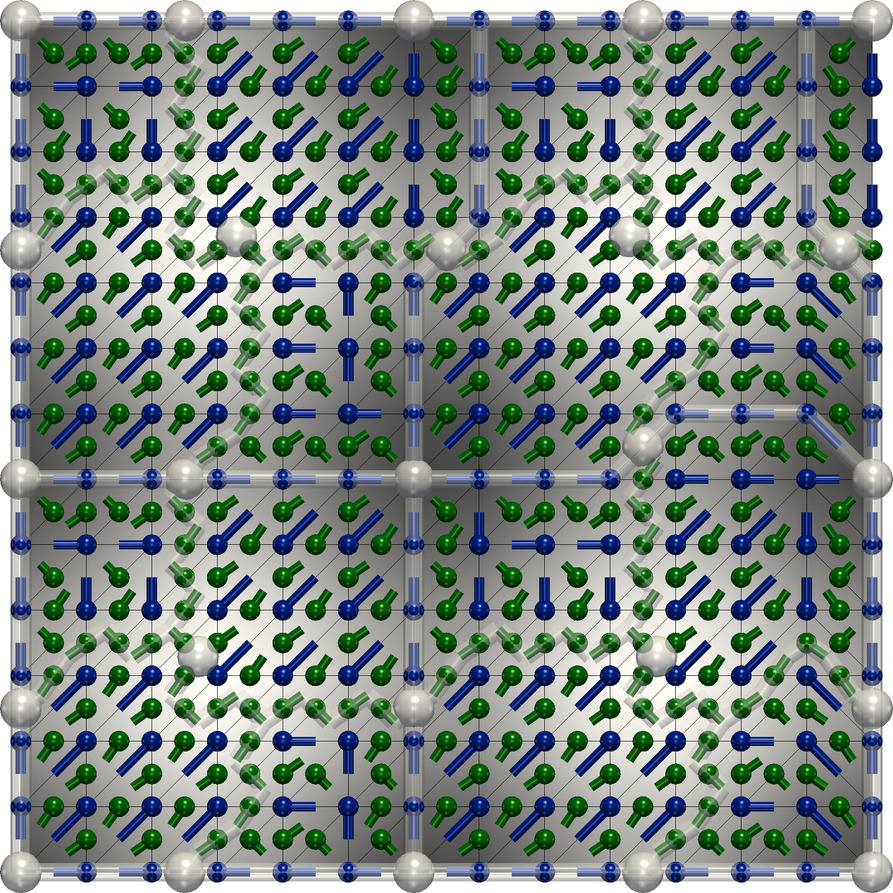}
  \hfill
  \includegraphics[width=0.32\linewidth]{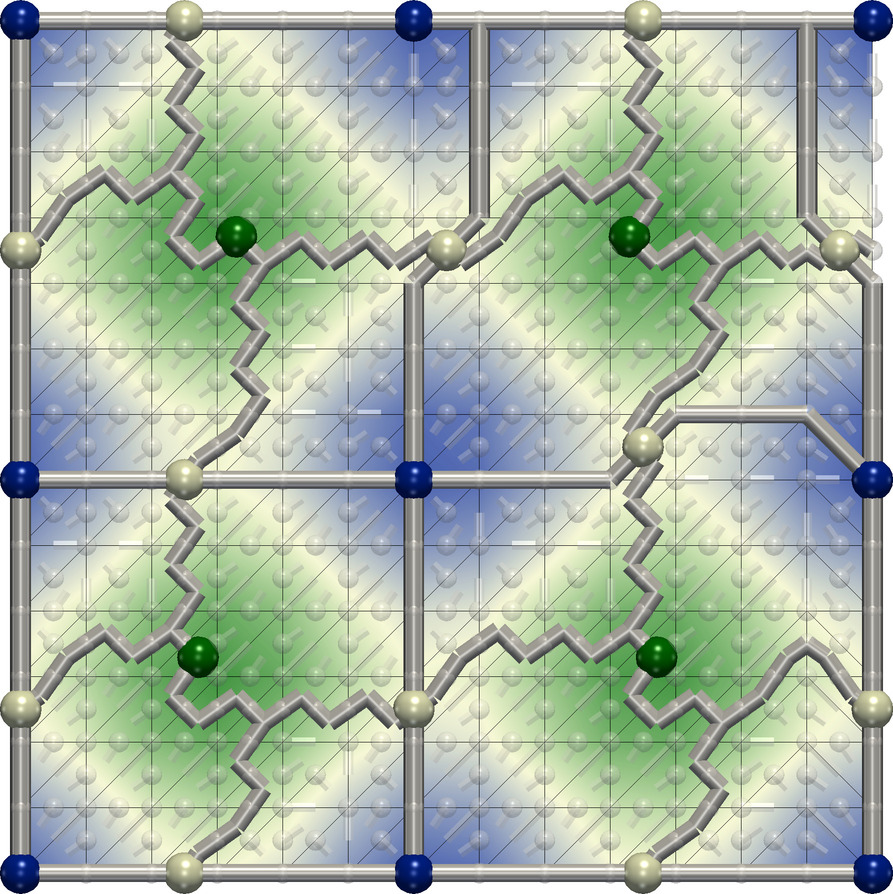}
\vspace{2.5ex}
  \imageCaption{Discrete Morse Theory on a 2D toy example. Left: input PL 
scalar 
field 
$f$. Center: discrete gradient (the origin of each pair is displayed with a 
sphere, vertex-edge and edge-triangle pairs 
are shown in 
blue and green). Right: discrete Morse-Smale complex 
$\morseSmale(f)$ obtained by extracting the $V$-paths emanating from critical 
simplices.
\vspace{-2ex}}
  \label{fig_discreteGradient}
\end{figure}

A \emph{discrete Morse function} is a function that assigns a scalar value to 
every simplex in $\domain$, such that 
each $i$-simplex $\simplex_{i} \in 
\domain$
has at most one
co-face $\simplex_{i+1}$ (resp. face 
$\simplex_{i-1}$) with lower (resp. higher) function value: 
$|\{ \simplex_{i+1} > \simplex_{i} ~ | ~ f(\simplex_{i+1}) \leq 
f(\simplex_{i}) \}| \leq 1$ and $|\{ 
\simplex_{i-1} < \simplex_{i} ~ | ~ f(\simplex_{i-1}) \geq f(\simplex_{i}) 
\}| \leq 
1$. Simplices for which these two numbers are zero are called \emph{critical 
simplices} and their dimension matches their index $\Index$. 
A \emph{discrete vector} is a pair of simplices $\{ \simplex_i < 
\simplex_{i+1}\}$.
A \emph{discrete vector 
field}  $V$ is a collection of such pairs such that each simplex 
appears in at most one pair. Then, a \emph{$V$-path} is a sequence of pairs of 
$V$
$\{ \simplex_i^0 < \simplex_{i+1}^0\}, \{ \simplex^1_i < \simplex^1_{i+1}\}, 
\dots, \{ \simplex^r_i < \simplex^r_{i+1}\}$ such that  $\simplex_i^j \neq 
\simplex_i^{j+1} < \simplex_{i+1}^j$ for each $j = 0, \dots, 
r$. A \emph{discrete gradient} is a discrete vector field for which all 
$V$-paths are monotonic and loop free. For these, $V$-paths are discrete 
analogs of integral lines, and simplices which are left unpaired in $V$ are 
critical. 
Then, the 1-dimensional cells of $\morseSmale(f)$  connected to maxima (resp. 
minima), called $1$\emph{-separatrices}, can be constructed by extracting 
ascending (resp. descending) $V$-path(s)
from $(d-1)$-saddles (resp. $1$-saddles). In 3D, its 2-dimensional cells, 
called $2$\emph{-separatrices} can be constructed with a descending (resp. 
ascending) breadth-first search traversal from $2$-saddles (resp. $1$-saddles) 
on the primal (resp. dual) of $\domain$. Last, the 1-dimensional cells of 
$\morseSmale(f)$ linking a 1-saddle $s_1$ to a 2-saddle $s_2$, called 
\emph{saddle-connectors}, can be extracted by computing the $V$-path(s), 
restricted to a given $2$-separatrix, linking $s_1$ to $s_2$.
\figref{fig_discreteGradient} illustrates these notions. 
Note that 
all saddles are non-degenerate and the 
transversal intersection  is respected by construction.

\shrinkedSection{Unified topological analysis and simplification}
\label{sec_topologicalSimplification}

Although it allows  a robust and consistent computation of 
$\morseSmale(f)$,
the 
DMT formalism is not directly compatible with the PL setting. 
DMT critical points are typically much more numerous than PL critical points in 
practice.
Moreover, in contrast to PL critical points which are located on vertices only, 
they are located on simplices of all dimensions.
This incompatibility
challenges the design of a unified framework such as TTK, for which the 
support 
of advanced analysis, robustly combining several topological abstractions,
can be highly valuable 
in practice.
Moreover, each topological abstraction traditionally comes with its own 
simplification mechanism.
This  
further challenges the development of a unified framework, as 
multiple instances of simplification algorithms 
(not necessarily consistent with each other) need to be implemented 
and maintained.
In this section, we describe an approach that addresses these two issues 
(unified representations and simplifications). 

\shrinkedSubSection{Initial discrete gradient}
Given an input function $f$ valued on the vertices of $\domain$, several 
algorithms \cite{gyulassy_vis08, robins11, ShivashankarN12, gyulassy_vis12, 
gyulassy_vis14} have been proposed for the construction of a discrete gradient 
(\secref{sec_dmt}). Among them, we 
focus on a variation of the 
elegant 
algorithm by 
 Shivashankar 
and Natarajan \cite{ShivashankarN12} for our initial 
gradient computation, as its localized nature will ease the following discussion 
and allows for a trivial and efficient parallelization. For completeness, we 
briefly sketch its main steps in \algoref{algo_initialDicreteGradient}.

The algorithm takes as an input a scalar field $f$ and an injective offset 
field $\offset$ (\secref{sec_criticalPoints}). It
visits the simplices of $\domain$ dimension-by-dimension.
For each dimension $i$, $i$-simplices are processed 
independently (and thus in parallel). The candidate set $C(\simplex_i)$ (line 
\ref{algo_candidate}) is constructed as the set of co-faces of $\sigma_i$ for 
which $\sigma_i$ maximizes the $i$-dimensional faces. 
$\simplex_i$ is then paired with the minimizer of this 
candidate set (line \ref{algo_minimizer}).
Here, the notions of minimizer and maximizer require the definition of a 
comparator (lines \ref{algo_startComperator} to \ref{algo_endComperator}), 
which iteratively compares two $i$-simplices $\simplex_i$ and $\simplex'_i$ by 
comparing their maximum vertices.
If 
these two vertices are identical ($\offset(v[j]) == \offset(v'[j])$), the 
next couple of maximum vertices will be considered until all vertices are 
visited.

\newtheorem{property}{Property}
\label{sec_matching}
We now discuss a key observation regarding
the above algorithm
\cite{ShivashankarN12}, 
which is useful  
for the development of a unified TDA framework.

\vspace{-1.5ex}
\begin{property}[PL Matching]
\label{prop_matching}
  Let $f$ be a PL scalar field defined on a closed, PL
  2 or 3-manifold $\domain$. Algorithm \ref{algo_initialDicreteGradient} will 
produce a discrete gradient field $V$ such that each PL critical point $p$ of 
index 
$\Index(p)$ will admit at least one critical simplex of dimension $\Index(p)$ in 
its star $\st{p}$.
\end{property}
\vspace{-1.5ex}

\begin{figure}[t]
  \includegraphics[width=\linewidth]{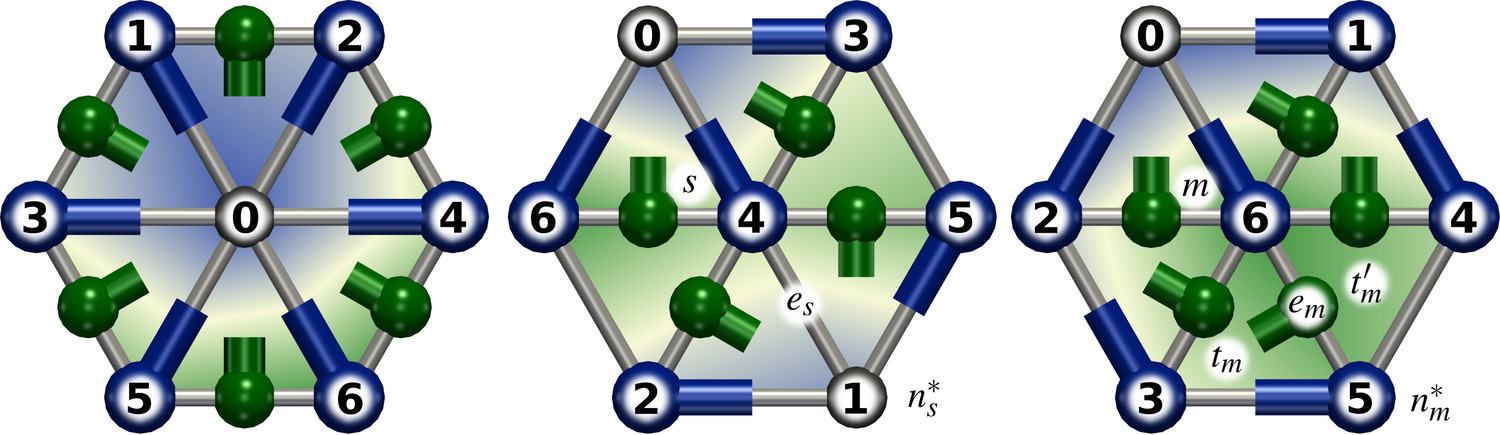}
  \vspace{-0.5ex}
  \imageCaption{
  \algoref{algo_initialDicreteGradient} leaves at least one vertex (0, 
left), one edge ($e_s$, center) and one triangle ($t'_m$, right) unpaired in 
the star of a 2D PL minimum (left), saddle (center), and maximum (right).
Vertex-edge and edge-triangle pairs 
are shown with blue and green balls-and-sticks respectively.
}
  \label{fig_2Dmatching}
\end{figure}

\newlength{\textfloatsepsave} 
\setlength{\textfloatsepsave}{\textfloatsep} 
\setlength{\textfloatsep}{-1ex}
\newcommand{\algosize}{\fontsize{6}{6.25}\selectfont}

\begin{algorithm}[t]
\algosize{
\SetKwInOut{Input}{input}
\SetKwInOut{Output}{output}
\SetKwFunction{Fn}{Function}{}{}
\Input{~ PL scalar field $f: \domain \rightarrow \mathbb{R}$ (with
injective offset field $\offset : \domain^0 \rightarrow \mathbb{R}$);}
\Output{~ Discrete gradient field $V$.}
\BlankLine

\Fn{$\simplex_i < \simplex'_i$}{
\nllabel{algo_startComperator}

$\quad v =$ \texttt{sort}$_{f, \offset}(\simplex{_i}^0)$\;
$\quad v' = $  \texttt{sort}$_{f, \offset}(\simplex'{_i}^0)$\;
$\quad$\For{$j \leftarrow i+1$ \KwTo $0$}{
  
  \lIf{$f(v[j]) < f(v'[j])$}{\Return true;}
  
  \lIf{$f(v[j]) == f(v'[j])  ~ \&\& ~ \offset(v[j]) < \offset(v'[j])$}{
    \Return true;
  }
  
  \lIf{$\offset(v[j]) \neq \offset(v'[j])$}{
    \Return false;
  }
}
$\quad$\Return false;
\nllabel{algo_endComperator}
}

\BlankLine

\Begin{

  \For{$i \leftarrow 0$  \KwTo $(d - 1)$}{
    \ForEach{$\simplex_i \in \domain$}{
      $C(\simplex_i) = \{\simplex_{i+1} > \simplex_i ~ | ~ \simplex_i = 
argmax_{\simplex'_i < \simplex_{i + 1}}f(\simplex'_i)\}$\; 
      \nllabel{algo_candidate}
      \If{$C(\simplex_i) \neq \emptyset$}{
        $V = V \cup \{\simplex_i < argmin_{C(\simplex_i)}f(\sigma_{i + 1})\}$\;
        \nllabel{algo_minimizer}
      }
    }
  }

}
\caption{Initial discrete gradient construction \cite{ShivashankarN12}}
\label{algo_initialDicreteGradient}
}
\end{algorithm}

This property can be justified based on two key observations \julien{(see 
\cite{shivashankar14} for an alternate discussion)}. First, \emph{(i)} 
If a $(d-1)$-simplex $\simplex_{d-1}$ maximizes both its $d$-co-faces, 
one of the two will be critical
as $\simplex_{d-1}$ can be paired only once. 
\emph{(ii)} 
For any $i$-simplex $\simplex_i$, its maximizing face is the one
which does not contain its minimizing vertex and no other face of 
$\simplex_i$ can be paired with it.
These two observations allow a simple and non-ambiguous
characterization of the pairing resulting from 
\algoref{algo_initialDicreteGradient} in the star of each PL critical point.

In 2D, critical simplices of dimension $0$ precisely coincide with PL minima 
(\figref{fig_2Dmatching}, left), 
since a local minimum is the maximizer of none of its incoming edges 
(observation \emph{(ii)}),
and will 
therefore never be paired.
By definition, the lower link $\lkminus{s}$ of a PL saddle $s$ is made 
of at least two connected components, 
each of which containing a local minimizing vertex.
Let $n_s^*$ be the second lowest vertex of this set of local minimizers 
(\figref{fig_2Dmatching}, center). $s$ will be paired by 
\algoref{algo_initialDicreteGradient} with the 
edge linking it to the lowest vertex of this set (line \ref{algo_minimizer}).
Let $e_s$ be the edge linking $s$ to $n_s^*$. 
$n_s^*$ cannot be paired with $e_s$ as $n_s^*$ is its minimizing vertex 
(observation 
\emph{(ii)}). 
The direct neighbors of $n_s^*$ on $\lk{s}$ are
necessarily higher than $n_s^*$ by definition. 
Thus, the co-faces of $e_s$ cannot be paired with it, as $e_s$ contains 
$n_s^*$, which is the minimizing vertex of both triangles (observation 
\emph{(ii)}).
Thus, the edge $e_s$ will not be paired by 
\algoref{algo_initialDicreteGradient} and will 
be critical.
Let $n_m^*$ be the global maximum of $f$ restricted to the link $\lk{m}$ of a 
PL maximum $m$ (\figref{fig_2Dmatching}, right). 
Let $t_m$ and $t'_m$ be the two triangles of $\st{m}$ connected 
to the edge $e_m$ linking $m$ to $n_m^*$. $e_m$ is by definition the maximizer 
of both triangles $t_m$ and $t'_m$ but can be paired with only one of them 
(observation \emph{(i)}).
Thus, the remaining triangle will be 
left unpaired by \algoref{algo_initialDicreteGradient}, and thus critical.
A similar, yet slightly more involved, 
reasoning applies constructively in 3D, 
as described in Appendix A (supplemental material).

\shrinkedSubSection{PL-compliant discrete gradient}
The \emph{PL matching} property (\secref{sec_matching}) indicates that we are 
guaranteed to find one DMT critical simplex for each PL critical point. This 
allows us to introduce an injective map $\xi : PL(f) \rightarrow DMT(f)$, 
from the 
set of PL critical points $PL(f)$ to that of DMT critical simplices $DMT(f)$, 
that maps each 
PL critical point $p$ of index $\Index(p)$ to a unique DMT critical simplex  of 
dimension $\Index(p)$  in its star. If multiple DMT critical 
$\Index(p)$-simplices exist in $\st{p}$, we select the highest 
as 
$\xi(p)$. 
We relax $\xi$ in the presence of a degenerate saddle $s$
and 
allow $s$ to be matched with $m > 1$ DMT critical simplices, where $m$ is 
the 
multiplicity of $s$.

However, in practice, the majority of the simplices of $DMT(f)$ will be left 
unmatched by $\xi$, requiring an additional cleanup procedure, which we 
introduce here.
Let $DMT'(f)$ be this set:
$DMT'(f) = \{ \simplex \in DMT(f) ~ | ~ \xi^{-1}(\sigma) = 
\emptyset \}$.
These simplices are extraneous singularities which 
can be considered as artifacts of the discrete gradient construction. 
Thus, we introduce  a procedure to simplify the gradient 
field $V$, such that the simplices of $DMT'(f)$ are no longer critical.

\setlength{\textfloatsep}{\textfloatsepsave}

\begin{figure}[b]
  \vspace{-1ex}
  \includegraphics[width=\linewidth]{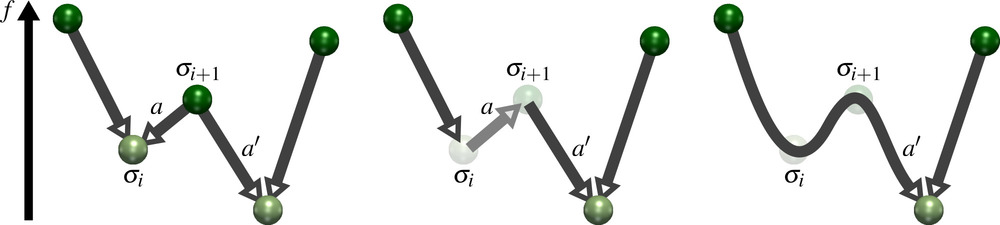}
  \vspace{-4.5ex}
  \caption{Removing a critical simplex pair from $DMT'(f)$. The arc $a 
\in G_0$ which minimizes its function difference is selected (left). Its 
corresponding $V$-path is reversed (center). The connectivity of $G_0$ is 
updated (right).\vspace{-3.5ex}}
  \label{fig_graphSimplification}
\end{figure}

\begin{figure*}
\vspace{-1ex}
  \centering
  \includegraphics[height=3.8cm]{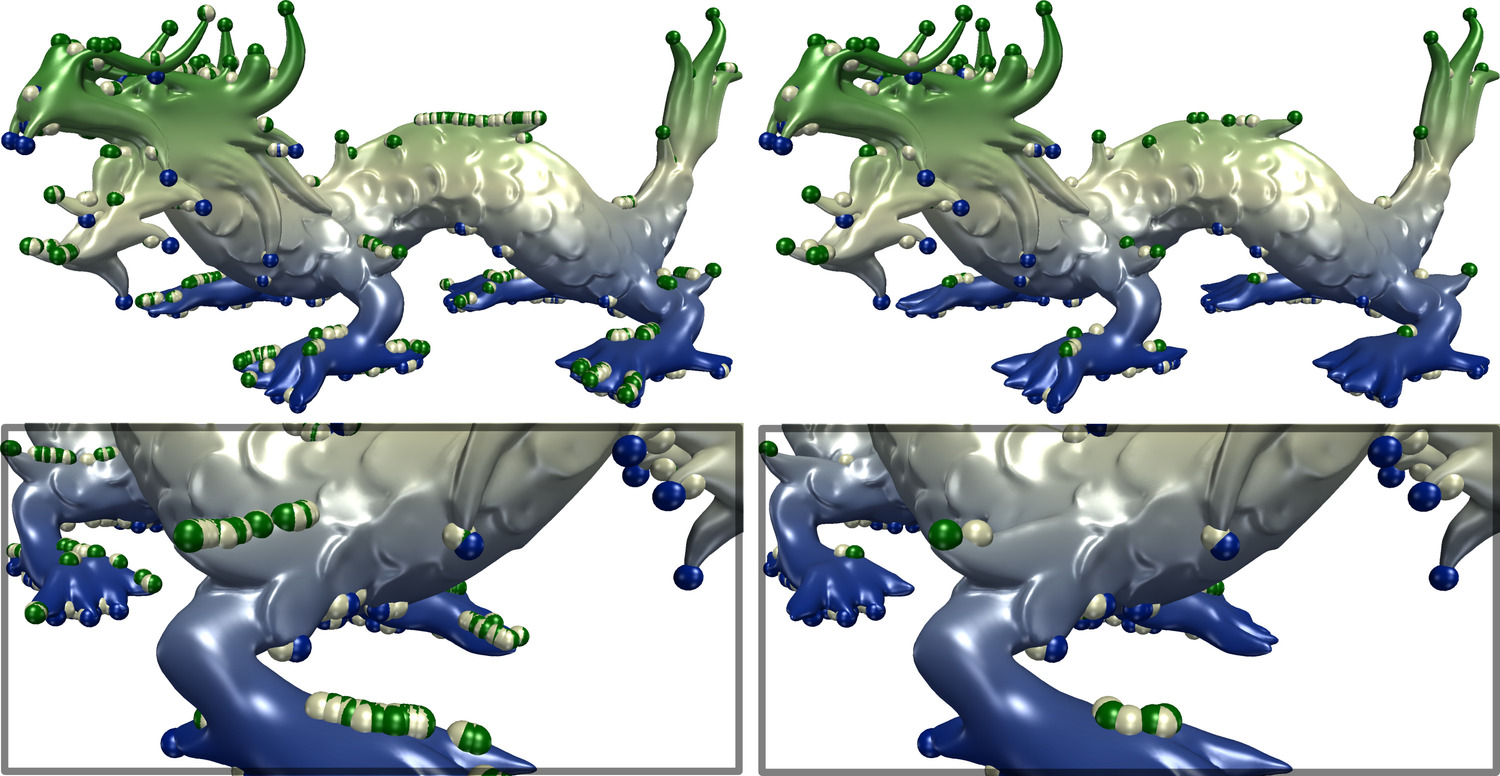}
  \hfill
  \includegraphics[height=3.8cm]{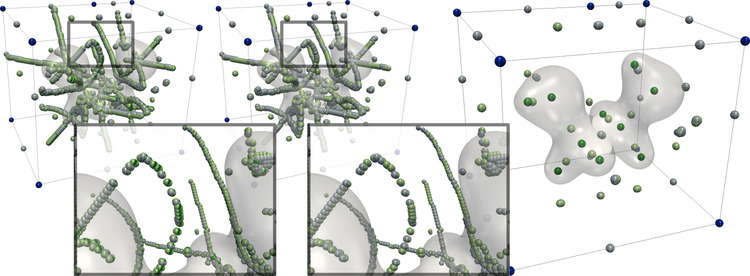}
  \vspace{-1.75ex}  
  \caption{
Critical simplices (spheres) before  and after 
PL-compliance on a 2D 
(left, 140k triangles) and 3D example (right, 10.5M tetrahedra). 
Each critical simplex is colored according to its dimension, 
from 
dark blue (minima) to dark green (maxima). 
In 3D, $|DMT(f)|$  decreases from 6,109 (left) to 2,997 
(center) after the removal of non PL saddle-maximum pairs, and to 93 (right)
after the removal of the non PL saddle-saddle pairs.}
\vspace{-2.75ex}
  \label{fig_plCompliantResult}
\end{figure*}

\noindent
\textbf{Saddle-maximum pair removal:}
Discrete Morse Theory \cite{forman98} 
indicates that two critical simplices $\simplex_i$ and $\simplex_{i+1}$
linked by a unique $V$-path $P$ can be simplified
by reversing the gradient precisely along $P$. Such an operation 
guarantees that $V$ indeed remains loop-free, that $\simplex_i$ and 
$\simplex_{i+1}$ are no longer critical and that no other simplex becomes 
critical. We make use of this property in the following (see 
\figref{fig_graphSimplification}). First, we construct 
the graph $G_0$, whose nodes correspond to the critical $d$ and 
$(d-1)$-simplices of $DMT'(f)$ and whose arcs correspond to $V$-paths linking 
them. $G_0$ is the subset of the $1$-separatrices of $\morseSmale(f)$ ending in 
maxima to be removed. Next, we visit each arc $a = (\simplex_i, \simplex_{i+1}) 
\in G_0$ in increasing order of function value difference.
If  $a$ 
is the only $V$-path connecting $\simplex_i$ to 
$\simplex_{i+1}$, we remove $a$ from $G_0$, reverse its $V$-path, 
and update the connectivity of $G_0$:
each arc $a' 
\in G_0$ that was connected to $\simplex_{i+1}$ gets reconnected to the 
$(i+1)$-simplices of $G_0$ that were connected to $\simplex_i$ and the function 
value difference for $a'$ and the corresponding $V$-path are updated. 
All other connections to $\simplex_i$ are removed. This process continues 
iteratively to remove
all the $d$-simplices of $G_0$.


\noindent
\textbf{Saddle-saddle pair removal: }
For PL $3$-manifolds, an extra step is 
required to remove saddles of $DMT'(f)$ that are connected in pairs by 
$V$-paths. Similarly to $G_0$, we construct the graph $G_1$ whose 
nodes correspond to the remaining $1$ and $2$-saddles of $DMT'(f)$ and whose 
arcs correspond to $V$-paths linking them. $G_1$ is the subset of the 
saddle-saddle connectors of $\morseSmale(f)$ linking $1$ and 
$2$-saddles which remain to remove. Next, $G_1$ is iteratively processed as 
described above for $G_0$.

For closed PL manifolds, the \emph{PL matching} property is guaranteed and our 
algorithm will remove all critical simplices of $DMT'(f)$. 
Thus, each remaining critical simplex of $V$ will be located in the star of a 
PL critical point, and we say that $V$ is \emph{PL-compliant}.
Note that this drastically differs from the \emph{conforming}
algorithm \cite{gyulassy_vis14}, which constrains the 
separatrices of $\morseSmale(f)$ to match an input 
segmentation.
This also differs from the \emph{assignGradient2} procedure 
\cite{ShivashankarN12}, which only simplifies  critical simplices if 
they are 
adjacent to each other, although they can be arbitrarily far from each other in 
practice, hence requiring our algorithm for the complete removal of the 
simplices of $DMT'(f)$.
The impact of the boundary of $\domain$ on our algorithm is discussed in 
\secref{sec_discussion}.


\shrinkedSubSection{Unified topological simplification}
Now that $V$ is PL-compliant,
it is possible to robustly combine multiple PL topological abstractions with 
the discrete Morse-Smale complex in a single pipeline (Figs \ref{fig_teaser} 
and \ref{fig_simplification}),
the exact 
correspondence between $DMT(f)$ and $PL(f)$ being given by $\xi$.
As a byproduct,
one can now 
mutualize
the topological simplification procedure 
traditionally used for multi-scale analysis and exploration, by pre-simplifying 
the data itself, prior to the computation of the topological abstraction under 
consideration. Several combinatorial algorithms \cite{edelsbrunner06, attali09, 
bauer11, tierny_vis12} have been proposed for this purpose. 
%
Here, we focus on 
the approach by Tierny and Pascucci \cite{tierny_vis12} as it supports 
arbitrary simplification heuristics. In particular, given a list of extrema 
$E$ to maintain, this algorithm will minimally modify both $f$ and $\offset$
such that $PL(f)$ only admits the critical points of $E$ as extrema. 
\julien{This is achieved by an iterative flattening of the sub-level set 
components corresponding to the critical points to remove \cite{tierny_vis12}.}
The set of maintained extrema $E$ can be selected according to 
 persistence (\secref{sec_persistence}\julien{, in which case the output 
topological abstractions will be guaranteed to be consistent with 
post-process
simplification schemes \cite{pascucci07, gyulassy_vis08}}) or any 
application-driven 
metric. 
Once $f$ and $\offset$ have been pre-simplified, 
the topological abstractions under consideration can be constructed for this 
simplified data (\figref{fig_simplification}). Our PL-compliant discrete 
gradient algorithm will 
 guarantee that the discrete Morse-Smale complex complies to this 
simplification.

\begin{table}[b]
  \centering
  \vspace{-2ex}
  \tableCaption{Running time of the different steps of the PL-compliance 
algorithm
(in seconds, with 12 cores) for the examples of
\figref{fig_plCompliantResult}. $PL(f)$, S-M,  S-S, $\morseSmale(f)$ and
$\morseSmale'(f)$ respectively stand for the computation times for the PL 
critical points, the saddle-maximum pair removal, the saddle-saddle pair
removal, the total Morse-Smale complex construction (including gradient 
processing) with and without the 
PL-compliance.}
  \label{tab:runningTimes}
  \scalebox{0.675}{
    \centering
    \begin{tabular}{l|rr|r|rrr|rr}
    \toprule
    Dataset &$|DMT(f)|$ & $|PL(f)|$ &
    \algoref{algo_initialDicreteGradient} & 
$PL(f)$ & S-M & S-S &
    $\morseSmale(f)$ & $\morseSmale'(f)$ \\
    \midrule
    Dragon & 1,118 & 318 & 0.016& 0.018 & 0.004 & 0 & \textbf{0.074} & 0.072\\
    EthaneDiol & 6,109 & 93 &4.943  & 1.525 & 0.144 & 3.864 & \textbf{13.829} & 
11.804\\
    \bottomrule
    \end{tabular}
  }
\end{table}

\shrinkedSubSection{Performance}
The initial gradient computation (\algoref{algo_initialDicreteGradient}) and 
the PL critical point extraction (\secref{sec_criticalPoints}) are both 
linear-time algorithms, which we implemented in parallel with OpenMP. 
The simplification of 
$G_0$ and $G_1$ is implemented sequentially, 
in a similar way 
to previous Morse-Smale complex simplification approaches
\cite{gyulassy_vis08, ShivashankarN12}. 
The optional  
pre-simplification of the data \cite{tierny_vis12}
is implemented sequentially and each separatrix of the Morse-Smale complex 
(\secref{sec_dmt}) is extracted in parallel. \tabref{tab:runningTimes} reports 
performance numbers, obtained on a Xeon CPU (2.6 GHz, 2x6 cores), for the 
examples shown in \figref{fig_plCompliantResult}. This table 
shows that, when fully constructing $\morseSmale(f)$
(with all 
separatrices), the PL-compliance step results in a small overhead overall. Note 
that in practice, the extraction of the geometry of the saddle-saddle 
connectors (used for the removal of the corresponding pairs by path reversal) 
is more demanding than that of saddle-maximum separatrices (column \emph{S-M} 
and \emph{S-S} in \tabref{tab:runningTimes}), as it requires  computing the 
corresponding 2-dimensional separatrices.

\shrinkedSection{Cached triangulation data structure}
\label{sec_triangulation}

Combinatorial TDA algorithms mostly involve mesh traversal routines. 
Thus, 
they must build on top of time efficient data structures. This section 
describes a new triangulation data structure designed to optimize time 
efficiency, while adjusting its memory footprint on demand.

\shrinkedSubSection{Traversal specifications}
\label{sec_specification}
Two core types of traversal queries are typically found in TDA algorithms, 
which consist of either accessing the faces or the co-faces of 
a simplex, as illustrated in the 
following traversal examples.

\noindent
\textbf{(i) Boundary:} A frequent test in TDA algorithms consists of 
checking if an $i$-simplex $\simplex_i$ is located on the boundary of $\domain$. 
This can be achieved by querying the $(d-1)$-co-faces of $\simplex_i$ 
(where $d$ is the dimension of $\domain$) and verifying if some admit only one 
$d$-co-face. 

\noindent
\textbf{(ii) Skeleton:} It is often necessary in TDA  to access the 
$k$-skeleton of $\domain$ (all $i$-simplices of $\domain$, 
such that $i \leq k$). 
For instance, the construction of the merge tree 
(\secref{sec_preliminariesRG}) requires  accessing the $1$-skeleton of 
$\domain$. 


\noindent
\textbf{(iii) Link:}
The star and the link are key notions to characterize the neighborhood of a 
simplex (\secref{sec_preliminariesInput}).
In particular, the extraction of the critical points of 
$f$ (\secref{sec_criticalPoints}) requires constructing the link $\lk{v}$ of 
each vertex $v$, and to classify it into its lower $\lkminus{v}$ and upper 
$\lkplus{v}$ links. 
This construction can be achieved by 
querying the $d$-co-faces of $v$, then querying their $(d-1)$-faces (which do 
not intersect $v$, to obtain $\lk{v}$) and then querying their $0$-faces to 
examine if a vertex of $\lk{v}$ is lower or higher than $v$. 
A similar classification needs to be performed on the link  $\lk{e}$ of each 
edge $e$ for Jacobi set extraction \cite{tierny_vis16}.

\noindent
\textbf{(iv) Face / co-face:}
Another typical traversal example 
consists of querying for each $i$-simplex, its list of $(i-1)$-faces and 
$(i+1)$-co-faces. This 
query is typical 
for the computation of the 
discrete gradient (\algoref{algo_initialDicreteGradient}), where maximizing 
faces are first identified to construct a candidate set of co-faces  
(line \ref{algo_candidate}). Since this algorithm performs such 
queries for each dimension, the $(i-1)$-faces and $(i+1)$-co-faces of each 
$i$-simplex must be efficiently accessed.

As shown in the above examples, TDA algorithms may potentially need to 
access the $k$-faces and $l$-co-faces of each $i$-simplex, for \emph{arbitrary} 
$k$ and $l$ such that $0 \leq k < i < l \leq d$. 
Thus, we designed our triangulation data structure to support 
\emph{all} the following types of traversal operations:
\begin{enumerate}
  \item{\vspace{-1.5ex}Given a vertex $v$, access its $1$-, $2$-, 
and $3$-co-faces;}
  \item{\vspace{-1.5ex}Given an edge $e$, access its $0$-faces and its 
$2$- and $3$-co-faces;}
  \item{\vspace{-1.5ex}Given a triangle $t$, access its $0$- and $1$-faces and 
its 
$3$-co-faces;}
  \item{\vspace{-1.5ex}Given a tetrahedron $T$, access its $0$-, $1$-, and 
$2$-faces.\vspace{-1.5ex}}
\end{enumerate}

\shrinkedSubSection{Explicit triangulation}
\label{sec_explicit}
To optimize time efficiency, 
we 
designed our data structure to support the above traversal queries through 
constant time lookups. 
%
Given an input cell-based representation (\secref{sec_relatedTriangulation}) 
providing a list of 3D points $\pointList$ and a list $\simplexList$ of 
$d$-simplices 
(representing each $d$-simplex by a list of vertex identifiers), we construct 
our explicit triangulation data structure as follows. The exhaustive list of 
edges can be constructed by enumerating all the unique vertex identifier pairs 
present in $\simplexList$. This can be done in $O(|\simplex_d|)$ steps, where 
$|\simplex_d|$ is the number of $d$-simplices in $\domain$, by using a 
vertex-driven lookup table, which progressively stores the constructed vertex 
pairs on a per vertex basis, to avoid potential edge duplicates. Similarly, if 
$d$ equals 3, the exhaustive list of triangles can be constructed by enumerating 
all the unique vertex triplets present in $\simplexList$, which can also be 
done 
in $O(|\simplex_d|)$ steps by using a vertex driven lookup table to avoid 
potential triangle duplicates. 
Given these list of $1$-, $2$-, and $3$-simplices, one can simply iterate over 
the list of $k$-simplices ($k \in \{1, 2, 3\}$) to store for each vertex its 
list of $k$-co-faces. This is done in $O(|\simplex_k|)$ steps. The construction 
of such lists enables supporting the first type of traversal operation 
(\secref{sec_specification}) with constant time lookups.
Given the list of $k$-co-faces for each vertex, one can construct the list of 
co-faces for each edge. Given an edge $e = (v_0, v_1)$, the list of 
$k$-co-faces of $e$ can be obtained as the intersection of the sets of 
$k$-co-faces of $v_0$ and $v_1$. This can be done in $O(|\simplex_1|)$ steps. 
The list of co-faces for each triangle can be constructed similarly in 
$O(|\simplex_2|)$ steps. Last, given the list of $l$-co-faces of each 
$k$-simplex, the list of $k$-faces of each $l$-simplex is built in 
$O(|\simplex_l|)$ steps.
The pre-construction of all the lookup tables mentioned above allows for 
time efficient queries based on constant time lookups for all the traversal 
types described in \secref{sec_specification}, at the expense of an excessive
memory footprint.

To limit this footprint, we introduce a preconditioning mechanism. 
This mechanism exploits the fact that many of the lookup tables described 
above can be constructed independently. 
Our preconditioning 
mechanism enforces third-party algorithms to
clearly specify in advance (typically in an initialization step)
the type of traversal queries they are going to perform. For instance,
regarding 
the skeleton traversal (\secref{sec_specification}), to access the $1$-cofaces 
of 
each vertex $v$, a developer needs to call at initialization the 
\texttt{preprocessVertexEdges()} precondition function of our 
data structure, which will build, once for all and only if they have not been 
constructed yet, the list of $1$-simplices of $\domain$ and the list of 
$1$-co-faces for each vertex. 
Thus, each traversal query supported by our 
triangulation data structure is associated with its own precondition 
function to be called beforehand, which will only trigger the computation of 
the necessary lookup tables, if they have not been constructed yet. This 
preconditioning mechanism guarantees a memory footprint limited to the only 
required lookup tables.
This strategy is particularly 
useful when a single instance of triangulation 
is used by 
multiple algorithms, as it is typically the case in a dataflow model such as 
VTK's pipeline. 
There, our data structure 
will be progressively enhanced upon the precondition calls triggered by the 
 algorithms present in the pipeline (see Appendix B for 
implementation details).

\begin{figure}
  \centering
  \includegraphics[width=0.95\linewidth]{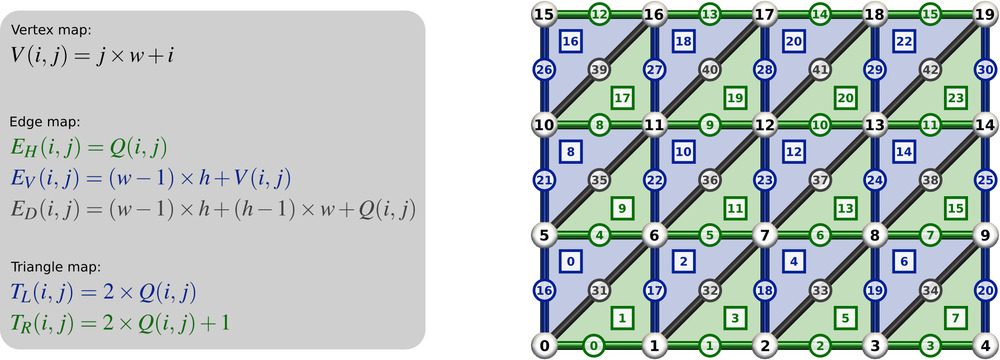}
\vspace{-2.5ex}
  \caption{
  Analytic expressions for the vertex, edge and triangle maps in a 
2D 
regular grid of width $w$ and height $h$. 
The identifier of each simplex is expressed as a function of the $(i, 
j)$-coordinates of its bottom left vertex.}
\vspace{-2.5ex}
\vspace{-1ex}
\label{fig_implicit}
\end{figure}

\shrinkedSubSection{Implicit triangulation}
\label{sec_implicitTriangulation}
In scientific visualization, the input scalar data is often 
provided as a 
regular grid. However, the typical size of these grids, especially in 3D, make 
the explicit storage of their triangulation impractical. To address this issue, 
we introduce an implicit triangulation data structure which emulates the above 
lookup tables 
in the case of regular grids.

In 2D, quads are considered to be pairs of triangles, whose common edge always 
follows the same diagonal. Given this regular structure, one can introduce 
analytic expressions to associate each $i$-simplex to a unique identifier. 
For 
brevity, we only discuss interior simplices here.
Let $V : \mathbb{N} \times \mathbb{N} \rightarrow \mathbb{N}$ be an injective 
function mapping a vertex to its identifier given its $(i, j)$-coordinates in 
a grid of width $w$ and height $h$ (in terms of number of vertices):
$V(i, j) = j \times w + i$. A similar function 
can be derived for each 
quad, given the $(i,j)$-coordinates of its bottom left vertex: $Q(i, j) = 
j\times(w -1) + i$.
One can introduce a similar convention, to uniquely enumerate edges, given
the $(i, j)$-coordinate of
their 
bottom left vertex, into horizontal ($E_H(i, j)$), 
vertical ($E_V(i, j)$) and diagonal ($E_D(i, j)$) edges (\figref{fig_implicit}).
%
%
Triangles can be enumerated similarly into left ($T_L(i, j)$) and right 
($T_R(i,j)$) triangles (\figref{fig_implicit}).

Once such a convention is established, the traversal queries of 
\secref{sec_specification} can be efficiently implemented by considering the 
pre-images of these functions. For instance, to perform traversal queries given 
an edge with identifier $e$,  the edge must first be classified  as 
horizontal ($e < (w - 1) \times h $) or vertical ($e < 
(w - 1) \times h + (h - 1) \times w $) or diagonal.
Then,
one can easily retrieve 
the $(i, j)$-coordinate of its bottom left vertex from the analytic 
expressions  $E_H$, $E_V$ or $E_D$. Then, the $0$-faces of $e$ are given by 
$V(i, j)$ for its bottom left vertex and $V(i + 1, j)$, $V(i, j+1)$ and $V(i + 
1, j+ 1)$ for its second vertex for horizontal, vertical and diagonal edges 
respectively. 
A very similar, yet more involved, strategy is derived in 3D, by considering 
that each voxel is composed of 6 tetrahedra. This regular structure also
allows introducing simplex maps analytically.
Then, given the identifier 
of an $i$-simplex $\simplex_i$, its $k$-faces and $l$-co-faces can be easily 
evaluated for arbitrary $k$ and $l$, by properly classifying $\simplex_i$ and 
considering the pre-image of the corresponding identifier function, as described 
above in 2D.
Since all identifiers are computed on the fly, this strategy 
effectively emulates our explicit data structure with no memory overhead. 
Note that 
in practice, we implemented both our explicit and implicit strategies 
with 
a common programming interface, which allows 
developers to manipulate our data structure generically, irrespective of 
its implicit or explicit nature.

\begin{table}[b]
  \centering
  \vspace{-2ex}
  \tableCaption{Running times (in seconds), memory footprint (MB) and overhead 
of 
our cached
triangulation data structure for the traversal examples of 
\secref{sec_specification} on 
the explicit (\emph{e-Time}, 1.25M 
tetrahedra, 71 MB in binary VTU) and the implicit (\emph{i-Time}) 
triangulation  of a $64^3$ grid.}
  \label{tab:runningTimesTraversal}
  \scalebox{0.65}{
    \centering
    \begin{tabular}{l||r|rr|rr||r}
    \toprule
    Traversal & Precondition & Memory & Memory & e-Time & Cache& i-Time \\
    Example & Time  & Footprint & Overhead & ~ & Speedup & ~\\
    \midrule
    Boundary Vertices ($0$)  & 1.082 & 48  & 68 \% & 0.001 & 997 & 0.003 \\
    Boundary Edges ($1$) &1.568 & 101 & 142 \% & 0.008 & 200 &  0.027\\
    Boundary Triangles ($2$) & 1.099& 50 & 70 \%& 0.011 & 100 & 0.038\\
    \midrule
    $1$-skeleton ($0$) & 0.200 & 23 & 32 \%& 0.025 & 9& 0.107\\
    \midrule
    Vertex Link ($0$) & 1.391 & 105 & 148 \% & 0.035& 41 & 0.135\\
    Edge Link ($1$) & 0.512 & 128 & 180 \% & 0.129 & 5 &1.857 \\
    Triangle Link ($2$) & 1.135 & 67 & 94 \% & 0.551 & 3 & 0.810\\
    \midrule
    Edge FaceCoFace ($1$)& 1.800 & 155 & 218 \% & 0.280 & 7 & 0.477\\
    Triangle FaceCoFace ($2$) & 1.310 & 119 & 168 \% & 0.447 & 4 & 0.451\\
    \bottomrule
    \end{tabular}
  }
\end{table}

\shrinkedSubSection{Performance}
\tabref{tab:runningTimesTraversal} reports performance numbers obtained on a 
Xeon CPU (2.6 GHz, 2x6 cores) with the explicit (1.25M tetrahedra, 71 MB in 
binary \texttt{VTU} file format) and implicit triangulations of a $64^3$ 
grid. Each test loops sequentially on the entire set of $i$-simplices 
of 
$\domain$ 
($i$ is reported in parentheses) and stores the result of the 
corresponding query.
This table shows that our explicit data structure indeed adapts its 
memory footprint depending on the type of traversal it undergoes.
In particular, the most memory-demanding traversals involve the list of 
edge co-faces ($i=1$) and the overhead required by our data structure ranges 
from 32\% to 218\%. Once it is preconditioned, our explicit  structure 
provides query times which are on par with its implicit counterpart. 
\tabref{tab:runningTimesTraversal} also reports the speedup of our 
explicit data structure, 
as the total time of a first 
run of each test (including preconditioning), divided by the time of a 
second run. Overall, this indicates that, if developers 
consent to multiply the memory footprint of their triangulation by a factor of 
up to $3.18$ in the worst case, typical time performance gains of one order of 
magnitude (and up to $3$) can be expected for their TDA algorithms.

\shrinkedSection{Software architecture}
\label{sec_software}

In this section, we describe our design decisions as well as the 
software engineering challenges that we faced when developing TTK.

\shrinkedSubSection{Design motivations}

The flexibility for software developers (challenge \emph{(ii)}, Sec. 1) 
required 
to write TTK with 
an efficient, low-level and portable programming language, such as 
C++. Moreover, to ease the integration of TTK into pre-existing, complex 
visualization systems, we designed its software architecture such that its  
implementation of TDA algorithms relies on no third party library. We achieved 
this by designing a first core layer  of \julien{dependency}-free, templated 
\emph{functor} classes (\secref{sec_architectureOverview}).

To make TTK easily accessible to end users (challenge \emph{(i)}), we tightly 
integrated it
into an advanced visualization programming environment, in particular VTK 
\cite{vtkbook}. Several alternatives to VTK could have been considered and we 
briefly motivate our choice here. In practice we found only a few 
downsides in using VTK, the most notable  being \julien{the modest time 
performance of}
their mesh data structures, which trade speed for generality. However, our 
cached triangulation data structure (\secref{sec_triangulation}) 
\julien{has been specifically designed to address}
this time efficiency issue. Apart from this downside, an integration into VTK 
comes with numerous advantages. First, VTK is easily accessible to the masses 
due to its open source and portable nature. 
Second, it provides a rich 
support for various standard file formats.
We believe this aspect is instrumental for end users' 
adoption. Third, it provides a rich, object-oriented rendering pipeline, 
which eases the development of interactive applications. Forth, 
 implementations following VTK's API  can easily be 
integrated into ParaView.
This 
integration does not only increase exposure to end users with a 
powerful visualization front end, but it also automatically provides,
without 
additional effort, software bindings for 
Python, which 
becomes more and more popular among engineers and scientists. Such a 
support is provided by ParaView's  \texttt{pvpython} Python wrapper.
Finally, a key argument 
in our choice 
towards VTK was its implementation of 
raw data pointers,
providing direct accesses to the primitive-typed (\texttt{int}, \texttt{float}, 
\texttt{double}, etc.), internal buffers of its data 
structures,
%
in reading and writing mode.
%
This capability is necessary to design a \julien{dependency}-free 
functor layer (\secref{sec_architectureOverview}) meant to interact with VTK, 
without data copy and without the inclusion of any VTK header.
Our last target, ease of extension for researchers (challenge \emph{(iii)}), is 
partly 
achieved due to our integration with VTK, as TDA developers  only need to
focus on the write-up of their core algorithm, without caring
about IO, rendering, or interaction, but still benefiting from the 
above advanced features offered by VTK and ParaView. We 
further improve the extension possibilities of TTK as described in 
\secref{sec_newModule}.

\begin{figure}
  \includegraphics[width=\linewidth]{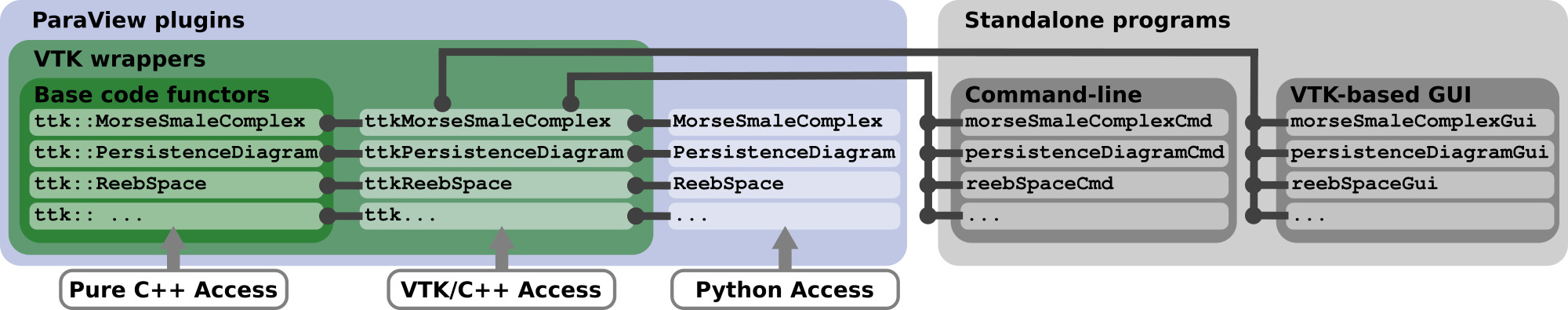}
  \vspace{-0.5ex}
  \imageCaption{TTK's software architecture is divided into 3 parts: the main 
library (green), ParaView plugins (blue) and automatically generated standalone 
programs (grey).  Developers can access TTK through each of its layer, either 
from \julien{dependency}-free C++, VTK/C++ or Python codes.}
\vspace{-2ex}
  \label{fig_architecture}
\end{figure}

\begin{figure*}
\vspace{-1ex}
\centering
  \includegraphics[width=0.3\linewidth]{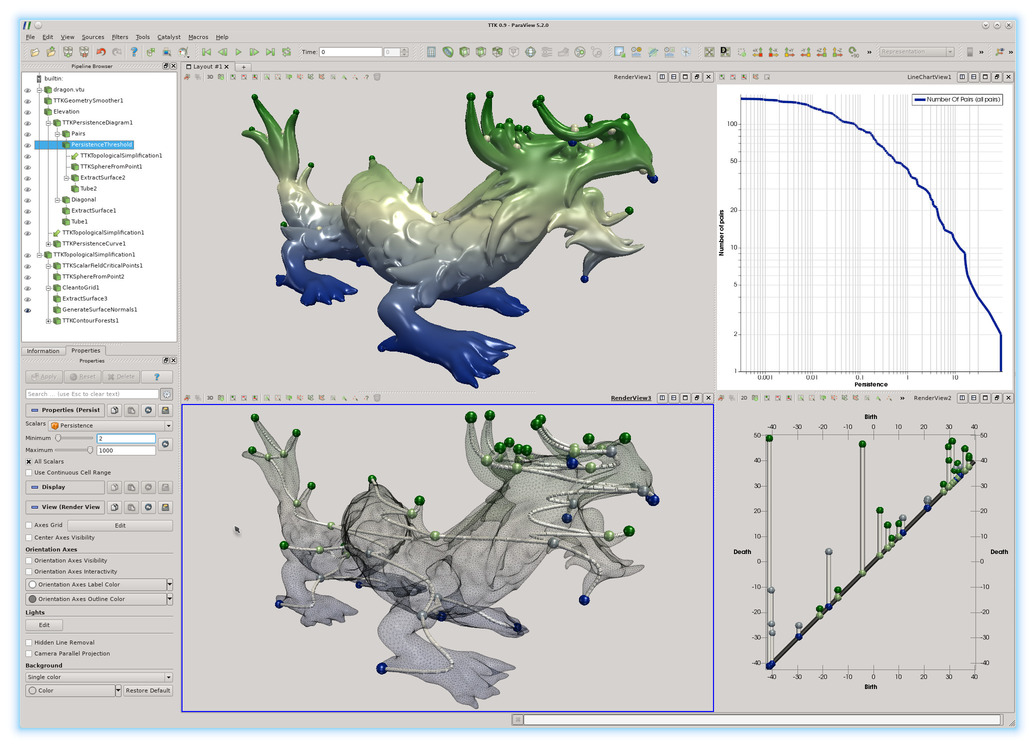}
  \hfill
  \includegraphics[width=0.3\linewidth]{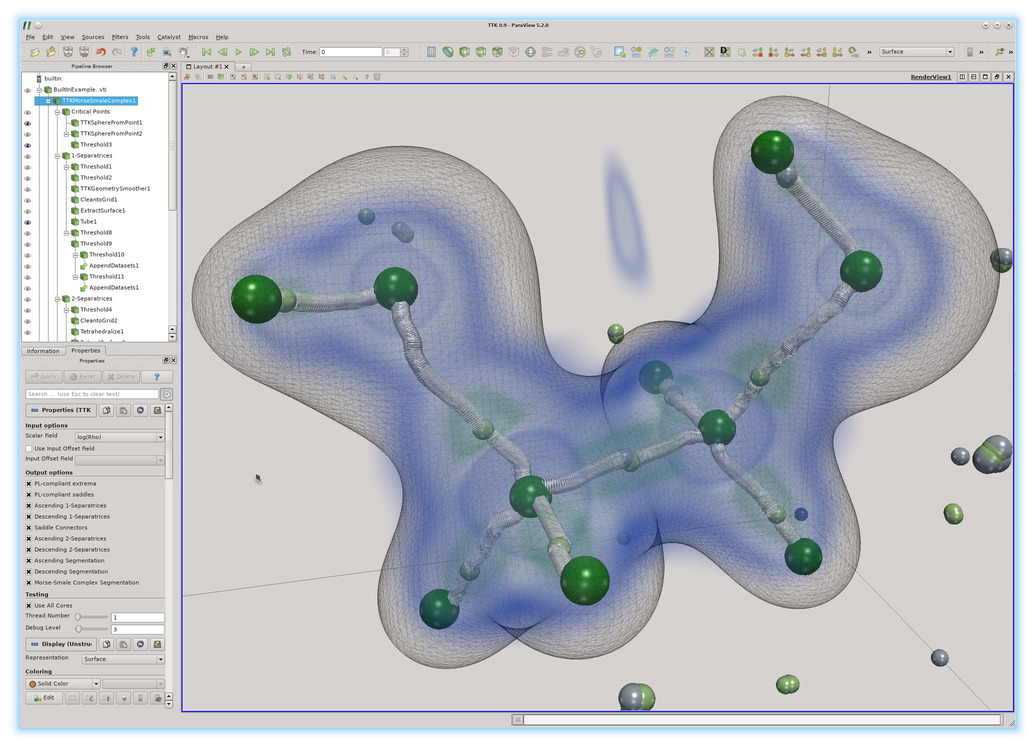}
  \hfill
  \includegraphics[width=0.3\linewidth]{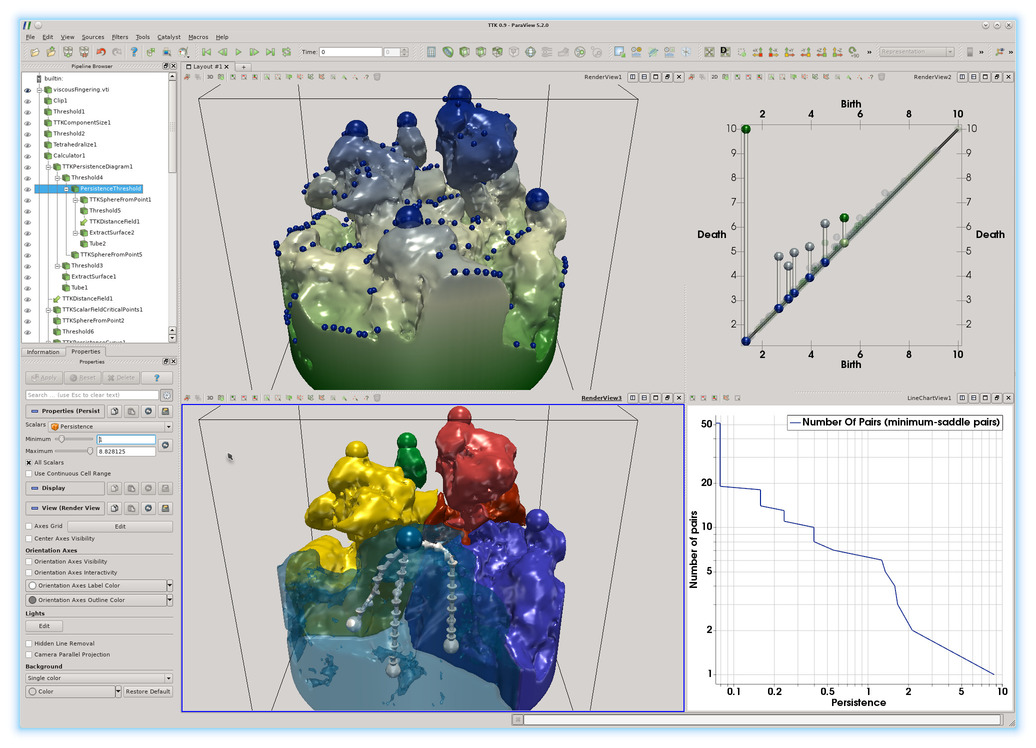}
  
  \includegraphics[width=0.3\linewidth]{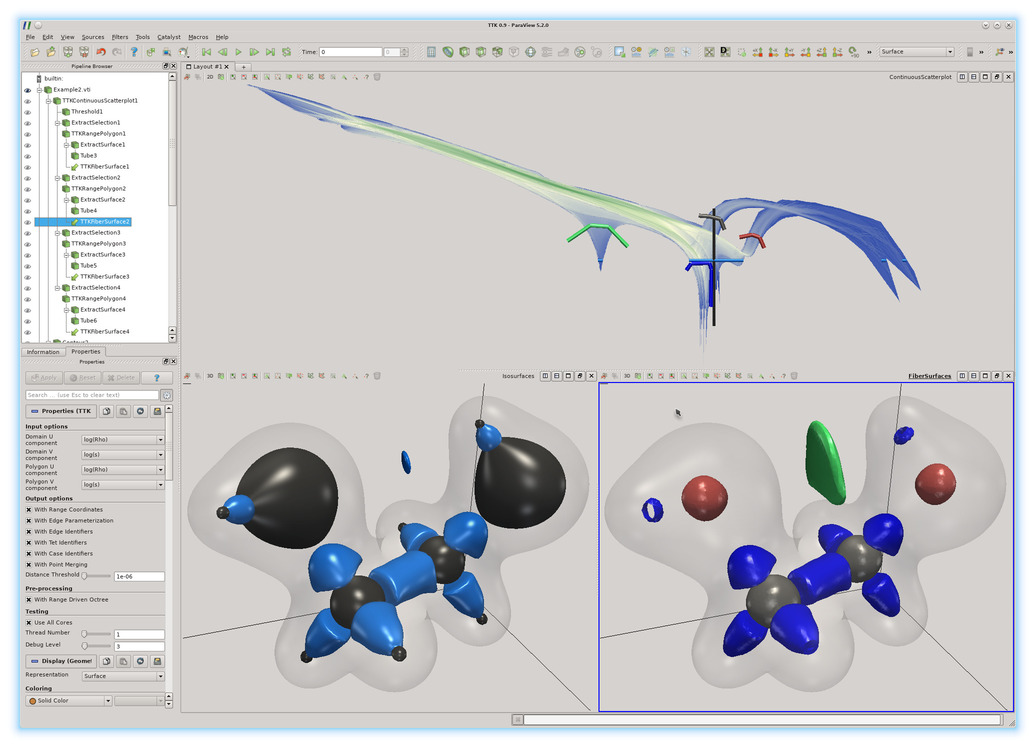}
  \hfill
  \includegraphics[width=0.3\linewidth]{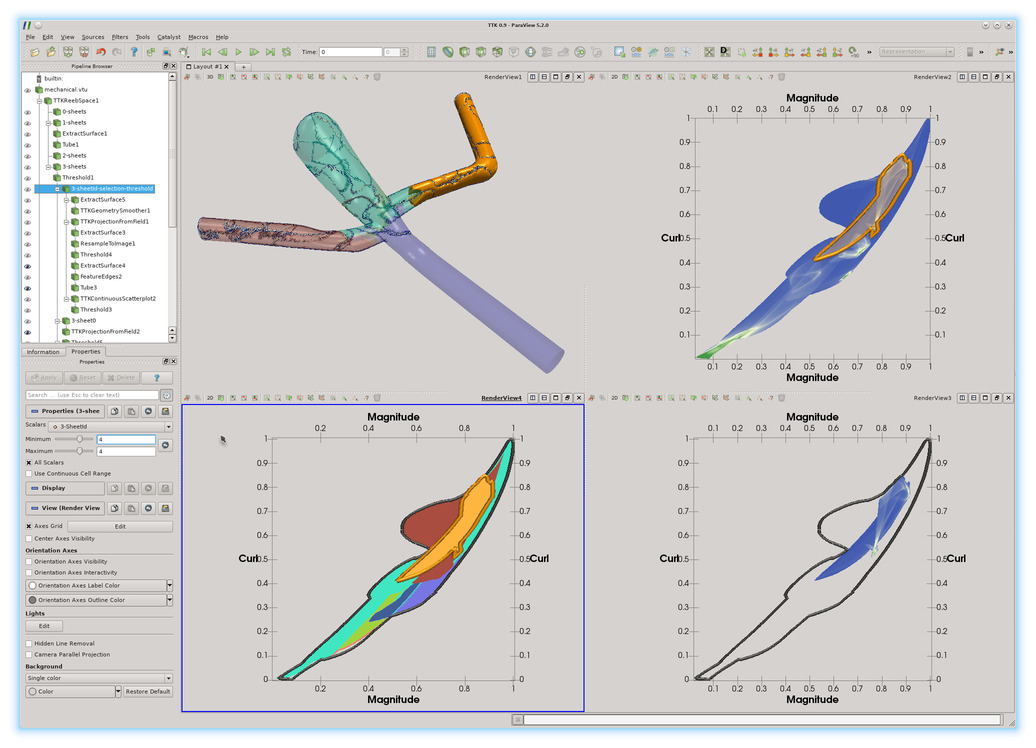}
  \hfill
  \includegraphics[width=0.3\linewidth]{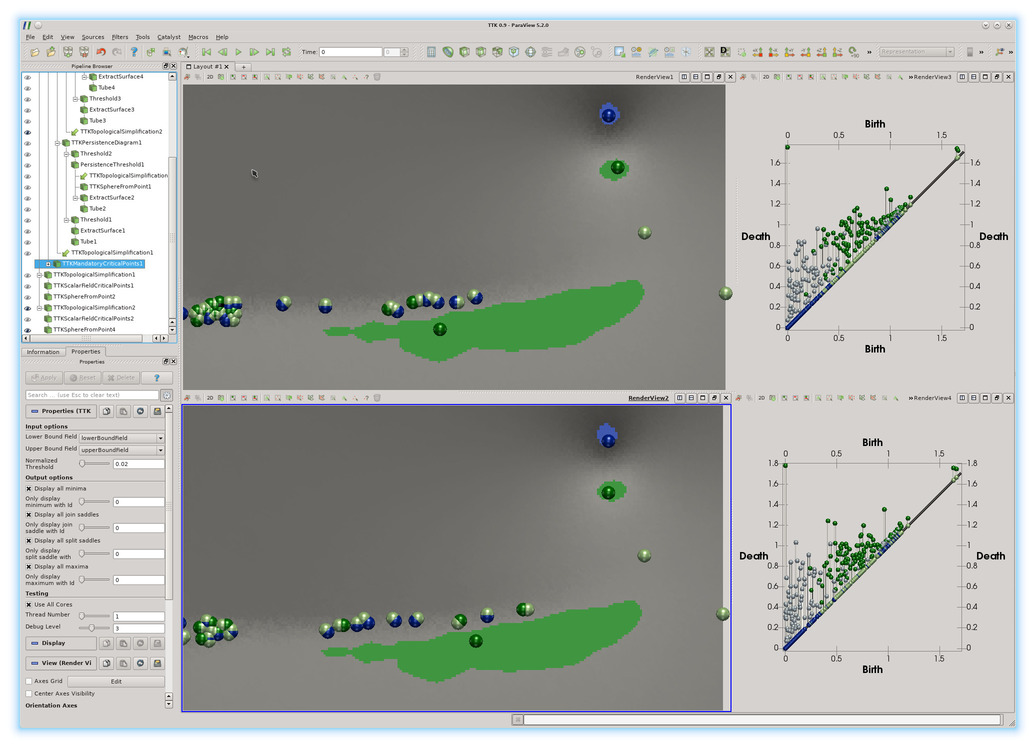}
  
  \vspace{2.5ex}
  \imageCaption{
  Gallery of TTK pipelines executed in ParaView. From top left to bottom right:
  \emph{(i)} The contour tree can be used for shape skeletonization.
Skeleton noise is removed by imposing the level of simplification dictated by 
the persistence diagram in the bottom linked view.
  \emph{(ii)} The interior $1$-separatrices of the Morse-Smale complex 
emanating from $2$-saddles directly capture the atomic and covalent structure 
of molecules (electron density).
  \emph{(iii)} The cells of the Morse-complex (colored regions) enable 
segmenting and tracking features during viscous fingering \cite{favelier16}.
  \emph{(iv)} Fiber surfaces (bottom right) from user strokes in the continuous 
scatter plot (top) enable an easy classification of the features in molecular 
systems \cite{klacansky16}.
  \emph{(v)} The Reeb space (top left) enables peeling continuous scatterplots 
(bottom left) into layers where the fibers are made of only one connected 
component in the volume \cite{tierny_vis16}. This enables localized inspections 
of the scatterplots (bottom right).
  \emph{(vi)} Mandatory critical points (colored regions) provide predictions 
on the location 
and function values 
of critical points 
for uncertain scalar fields with non-uniform error \cite{tierny_ev14}. These 
regions always admit at least one critical point for any function randomly 
generated from this error (top and bottom, vortices in 
computational fluid dynamics). 
}
  \label{fig_gallery}
\end{figure*}

\shrinkedSubSection{Overview}
\label{sec_architectureOverview}
TTK's software architecture is presented in \figref{fig_architecture}. 
In the following, we will 
refer to the running example of TTK's scalar field smoother, which 
iteratively smooths data by averaging neighbor values.

\noindent
\textbf{Base code functors} living in the \texttt{ttk} C++ namespace 
implement the TDA algorithms of TTK. They do not store data, but are given 
pointers to input and output buffers. Their processing routines are template 
functions, where the template parameters are the primitive data types used
by the input and output buffers. This template structure enables  writing
generic code, irrespective of the representation of the scalar data 
(\texttt{float}, \texttt{char}, etc.). 
These functors are also typically passed a pointer to a 
cached triangulation data structure (\secref{sec_triangulation}), which 
they precondition at initialization time, depending on the type of traversal 
they are going to perform. Note that these classes include no VTK header. For 
instance, 
the \texttt{ttk::ScalarFieldSmoother}
functor is given a pointer to the input and 
output data buffers and a pointer to a triangulation instance. It 
preconditions it at initialization with the function 
\texttt{preprocessVertexNeighbors()} and perform the smoothing based on the 
triangulation adjacency in a distinct template function.

\noindent
\textbf{VTK wrappers}
implemented as \texttt{vtkDataSetAlgorithm} filters \cite{vtkbook} connect 
each base code functor to VTK. In particular, these filters typically query a 
pointer to the internal buffers of the input and output objects and call the 
processing function of the corresponding functor with the appropriate template 
argument. 
For example, the \texttt{vtkScalarFieldSmoother}
selects the input field to 
smooth, allocates an output field and passes their pointers to 
its 
functor. Note that TTK functors can be used without VTK. An example is given 
in \figref{fig_teaser}.

\noindent
\textbf{ParaView plugins}
are automatically created from the VTK 
wrappers. The specification of each plugin is documented in an XML 
file, which is interpreted at build time. Such a file describes the 
options of the VTK wrappers which will be exposed to ParaView's 
GUI and Python binding. 
In our smoothing 
example, a developer would declare in the XML file the VTK wrapper 
function which controls the number of smoothing iterations. Then, the 
Python class automatically generated from the resulting ParaView plugin 
would provide a variable that enables tuning
this number.  
\figref{fig_teaser} illustrates some TTK plugins interacting together within a 
ParaView pipeline, as well as the corresponding Python script.

\noindent
\textbf{Standalone programs}
In certain situations, it may be desirable to run TDA algorithms in batch 
mode. Thus, we accompany each VTK wrapper with a generic command line program 
which 
reads all the datasets given as command line arguments, 
sends them to its VTK wrapper and writes all of its outputs to disk after 
execution. 
Developers only need to edit 
the main source file of this program to declare options 
(i.e. the number of smoothing iterations) to TTK's  command 
line parser, similarly to the ParaView XML specification. 
TTK also automatically 
accompanies each VTK wrapper with a VTK-based GUI, which 
behaves similarly to the command line program. Once opened, this GUI lets
users interact with the outputs of the VTK wrapper and, similarly to ParaView, 
let them toggle the display of each output with keystrokes.


\shrinkedSubSection{Implementing a new module for TTK}
\label{sec_newModule}
To ease the development of new TDA algorithms, TTK ships with Bash scripts 
which automate the creation and release of new TTK modules. In particular, the 
creation script generates a templated base code functor and its matching VTK 
wrapper, ParaView plugin, standalone command-line and GUI programs. At this 
point, each of these can be built and run. To implement their TDA algorithms, 
developers then only need to focus on the base code functor. The input and 
output specification should be enriched if needed from the default one in the 
VTK wrapper layer. Finally, options should be declared within the ParaView XML 
file and the standalone main source files. Also, another Bash script 
packages these components into a tarball for release purposes, with optional 
Doxygen online documentation and code anonymization.


\shrinkedSection{Software collection}
\label{sec_collection}
Each TTK module comes with its own base-code functor, VTK 
wrapper, ParaView plugin and command-line and VTK-GUI programs.

\noindent
\textbf{Scalar data:}
Critical points (\secref{sec_criticalPoints}) often correspond directly to 
features of interest in scalar data. TTK implements a combinatorial algorithm 
for their extraction in the PL setting \cite{banchoff70}. 
Merge trees and contour trees (\figref{fig_gallery}\emph{(i)}) are instrumental 
TDA 
abstractions for data segmentation tasks \cite{bremer_tvcg11, carr04, 
chemistry_vis14}. TTK implements their computation based on a multi-threaded 
approach \cite{gueunet_ldav16}.
The discrete Morse-smale complex (\secref{sec_dmt}), which is a key abstraction 
for data segmentation and for the extraction of filament structures 
\cite{chemistry_vis14, sousbie11, gyulassy07, shivashankar2016felix, favelier16}
(\figref{fig_gallery}\emph{(ii)}), is 
implemented with the PL-compliant approach described in 
\secref{sec_topologicalSimplification}.
Persistence diagram 
and curves (\secref{sec_topologicalSimplification}) help users appreciate the 
distribution of critical points and tune simplification thresholds. 
The extraction of the extremum-saddle and saddle-saddle pairs has been 
implemented as described in Secs.~\ref{sec_preliminariesRG} and \ref{sec_dmt}.
Topological simplification is implemented in an independent, unified manner 
(\secref{sec_topologicalSimplification}) for all 
the above abstractions by pre-simplifying the data with a combinatorial 
approach \cite{tierny_vis12}.
TTK also implements a few supporting features, 
including 
integral lines or data smoothing and normalization.

\noindent
\textbf{Bivariate scalar data:}
Jacobi sets
(\secref{sec_criticalPoints}) correspond to points where the volume folds onto 
itself when projected to the
plane by a bivariate function. TTK 
implements their combinatorial extraction 
\cite{tierny_vis16}. 
Reeb space (\secref{sec_preliminariesRG}) based
segmentation capabilities help users peel scatterplots views of the 
data (\figref{fig_gallery}\emph{(v)}). TTK implements  a recent combinatorial 
approach 
\cite{tierny_vis16}.
TTK also implements a few supporting features, including 
planar 
projections, continuous scatterplots \cite{bachthaler08}, 
fiber and fiber surfaces \cite{carr_ev15} based on user strokes 
\cite{klacansky16} (\figref{fig_gallery} \emph{(iv)}).

\noindent
\textbf{Uncertain scalar data:}
Uncertain scalar fields are becoming more and more prominent in applications. 
TTK implements a combinatorial approach for \julien{extracting}
\emph{mandatory 
critical points} \cite{tierny_ev14}, which predicts appearance regions for 
critical points, despite the uncertainty (\figref{fig_gallery}\emph{(vi)}). To 
support 
this, TTK also provides a module converting an ensemble dataset into 
a histogram-based representation of the data.

\noindent
\textbf{Miscellaneous:}
TTK also provides a number of support features, for data 
conversion, connected component size evaluation, geodesic distances, 
mesh subdivision, geometry smoothing, identifier fields, 
texture map computation from scalar data, or simplified sphere glyphs.

\noindent
\textbf{User experience:}
End users typically leverage ParaView's advanced interface capabilities to 
organize the linked views of their visualization (2D, 3D, line charts, 
etc.). TTK features are accessed through its plugins, used in 
conjunction with standard ParaView filters. 
For instance, in \figref{fig_teaser}(a), the user selected critical point pairs 
in  the persistence diagram (bottom right 2D linked view) by thresholding 
their persistence. Then, TTK's topological simplification was called to 
pre-simplify the data according to this  selection of critical points. 
Note 
that 
any 
other, application-driven, user selection can be used instead of 
persistence. At this point, any topological abstraction can be computed on this 
pre-simplified data, like the Morse-Smale complex. 
Next, the user can interactively modify the 
persistence threshold and 
all  the 
linked views are updated accordingly, as shown in our companion video. 
We refer the reader to TTK's website \cite{ttk} for more video tutorials.
Note that TTK's usage in ParaView requires no programming or scripting skill.

\shrinkedSection{Limitations and discussion}
\label{sec_discussion}
The PL matching property (\secref{sec_matching}) is only guaranteed for 
interior critical points. For non-closed domains, boundary PL 
critical points may admit no DMT critical simplex in their star. Thus, for 
simplicity, our implementation omits DMT critical simplices located on the 
boundary. 
This omission is not problematic in practice (\figref{fig_plCompliantResult}), 
although it may prevent the removal of certain pairs of DMT critical simplices, 
for which one or both simplices are located on the boundary.
\julien{More generally, although our cached triangulation data structure can 
support fairly general domains (such as non-manifold simplicial complexes), 
more research still needs to be done at the topological analysis level towards 
generality with regard to the input domain representation.}

The combinatorial procedure \cite{tierny_vis12} that TTK employs to 
pre-simplify scalar data only supports the removal of $(0, 1)$ and $((d-1), d)$ 
critical point pairs. More research needs to be done to find a practical 
algorithm for the removal of $(1, 2)$ critical point pairs, as 
homological simplification and reconstruction in $\mathbb{R}^3$ has been shown 
to be NP-hard \cite{attali13}.

Regular grids are treated by TTK as implicit triangulations (which is key to 
guarantee the PL matching property)
by considering the 6-tet subdivision of  each voxel. Thus, 
TDA algorithms which already natively  support arbitrary CW-complexes 
\cite{gyulassy_vis08, robins11, ShivashankarN12} may run more slowly with this 
representation as more cells will need to be considered.



We have been using preliminary versions of TTK internally for two years and 
published several research papers using it \cite{tierny_vis16, gueunet_ldav16, 
favelier16, vintescu17}. 
Five master-level students, with a C++
background  but no knowledge in rendering or user interfaces, have been 
using it on a daily basis. 
Typically, we found that TTK 
helped them shorten the prototyping phase of their research and 
access experimental validation faster,
some of these students starting 
implementations after just a week of training.
The pre-implemented IO and rendering support of ParaView was 
instrumental to provide an important time gain during prototyping.
Moreover, 
the ease offered by ParaView to combine
several plugins within a single pipeline 
also helped students easily leverage existing 
implementations in their prototypes to empower their own algorithms. 

\shrinkedSection{Conclusion}
\label{sec_conclusion}
This system paper presented the Topology ToolKit (TTK) \cite{ttk}, a software 
platform 
designed for topological data analysis (TDA) in scientific visualization. TTK 
has been designed to be easily accessible to end users (with ParaView plugins
and standalone programs)
and flexible for software developers 
with a variety of bindings. 
For researchers, TTK eases the prototyping phase of 
TDA algorithms, without compromise on genericity or time efficiency,
as developers only need to focus on the core routines of their algorithm; the 
IO, the rendering pipeline and the user interface capabilities being 
automatically generated. Although it focuses on TDA, we believe 
TTK also provides more generally an 
appealing infrastructure for any geometry-based visualization technique.

TTK builds on top of two main contributions: \emph{(i)} a unified
topological data representation and simplification and \emph{(ii)} a 
time-efficient triangulation data structure. Our PL-compliant discrete gradient 
algorithm allows 
to robustly and consistently combine 
multiple topological abstractions, defined in the 
Discrete or  PL settings, within a single 
coherent analysis, as showcased with the ParaView pipelines illustrated in this 
paper (Figs.~\ref{fig_teaser}, \ref{fig_simplification}, \ref{fig_gallery}). 
The genericity and time efficiency of TTK is in great part due to our 
novel cached triangulation data structure, which handles in a consistent 
way 2D or 3D explicit meshes or regular grids implicitly. In explicit mode, a 
preconditioning mechanism enables our data structure to deliver 
time-efficient traversal queries, 
while self-adjusting its memory footprint on demand.
For typical TDA traversals, we showed 
that if developers consent to reasonably
increase
the memory footprint of their 
triangulation, 
significant speedups
can be expected in practice for their algorithms. 
This aspect is particularly important when a single triangulation instance is 
accessed by multiple algorithms, as typically found in complex analysis 
pipelines.

TTK constitutes an invitation to a community-wide initiative to disseminate and 
benchmark TDA codes.
Although we did our best to implement in TTK a substantial  collection of TDA 
algorithms for scientific visualization, many more 
could be 
integrated in the future, including for instance Pareto set 
computation \cite{Huettenberger13} or robustness evaluation \cite{skraba16}.
%
Thus, we hope TTK will rapidly grow a developer community to help integrate 
more 
algorithms, as we specifically designed it to be easily extensible.
We also 
hope that future extensions of TTK will form the basis of an established 
software platform for TDA research code,
to improve 
the reproducibility and usability of TDA  in applications.
\acknowledgments{
\footnotesize{
This work is partially supported by the Bpifrance grant ``AVIDO'' (Programme
d'Investissements d'Avenir, reference P112017-2661376/DOS0021427) and by 
the National Science Foundation IIS-1654221.
We would like to thank Attila Gyulassy, Julien Jomier and Joachim Pouderoux for 
insightful discussions and Will Schroeder, who encouraged us to write this 
manuscript.}}

\clearpage

\newpage

\bibliographystyle{abbrv-doi}

\bibliography{paper}

\section*{Appendix A: PL-matching property in 3D}
Critical simplices of dimension $0$ still precisely coincide with PL 
minima (same argument as in 2D). Moreover, PL 1-saddles will admit at 
least one critical edge in their star (same argument as in 2D). 
By definition, the upper link $\lkplus{s_2}$ of a PL $2$-saddle $s_2$ is made 
of at least two connected components, each of which containing a local 
maximizing vertex. Thus, the restriction of $f$ on $\lk{s_2}$ will admit at 
least two PL maxima, and therefore at least one PL saddle $n_s$ (to satisfy the 
Morse inequalities), precisely separating $\lkplus{s_2}$  from 
$\lkminus{s_2}$. Due to the 2D argument, $n_s$ will admit a critical edge $e_s$ 
in its star on $\lk{s_2}$, linking it to a lower vertex $n'_s \in 
\lkminus{s_2}$.
Let $t_{s_2}$  be the triangle containing $e_s$ and $s_2$.
Since $e_s$ is not paired with its co-faces $t_s$ and $t'_s$, 
it means that $n'_s$ is the minimizing vertex of both triangles. Thus, $n'_s$ 
is also the minimizing vertex of the co-faces of $t_{s_2}$ in $\st{s_2}$. Thus, 
$t_{s_2}$ cannot be paired with its co-faces (observation \emph{(ii)}). Since 
$n'_s \in 
\lkminus{s_2}$, the edge $e_{s_2}$ linking $s_2$ to $n_s$ maximizes $t_{s_2}$. 
As discussed in the 2D case, $n_s$ has to be paired with the edge connecting it 
to its minimum neighbor on $\lk{s_2}$, $n''_s$.
Let  $t'_{s_2}$ be the triangle containing $s_2$, $n_s$ 
and $n''_s$.  $e_{s_2}$ maximizes both $t_{s_2}$ and  $t'_{s_2}$ but can be 
paired only once, with the one containing the 
minimizing vertex, $t'_{s_2}$. Thus, the triangle $t_{s_2}$ will be left 
unpaired by Alg. 1,
and thus critical.

%
%

Similarly, let $n_m^*$ be the highest vertex of the link $\lk{m}$ of a 3D PL 
maximum $m$. 
Then, $n_m^*$ is itself a 2D PL maximum on  $\lk{m}$.
Given the 2D argument, there must exist 
a triangle $t_m$ on $\lk{m}$  which contains $n_m^*$ and which is paired with 
no simplex of  $\lk{m}$. 
Moreover, $t_m$ also contains the vertex $n_m'^*$, 
which is the maximizer of the link of $n_m^*$ on that of $m$ (see 
the 2D argument). Let $n_m''^*$ be the remaining vertex of 
$t_m$.
Let $t'_m$ be the triangle of $\st{m}$ which contains both $n_m^*$ and $n_m'^*$ 
and let $T_m$ and $T'_m$ be its two adjacent tetrahedra in  $\st{m}$. We will 
consider that $T_m$ is the tetrahedron containing $n_m''^*$ and $T'_m$ that 
containing a fourth vertex $n_m'''^*$. Note that $n_m'''^*$ is lower than 
$n_m''^*$ (otherwise $t_m$ would not be unpaired on $\lk{m}$) and we have:
$f(n_m'''^*) < f(n_m''^*) < f(n_m'^*) < f(n_m^*) < f(m)$.
The triangle $t'_m$ is the maximizer of both $T_m$ and $T'_m$, 
however, 
it can be paired with only one of them (observation \emph{(i)}), its 
minimizing co-face $T'_m$, as it contains the lowest vertex, $n_m'''^*$, of the 
two tetrahedra.
Therefore, the tetrahedron $T_m$ will be left unpaired by 
Alg. 1,
and thus critical.

\begin{figure}
  \includegraphics[width=\linewidth]{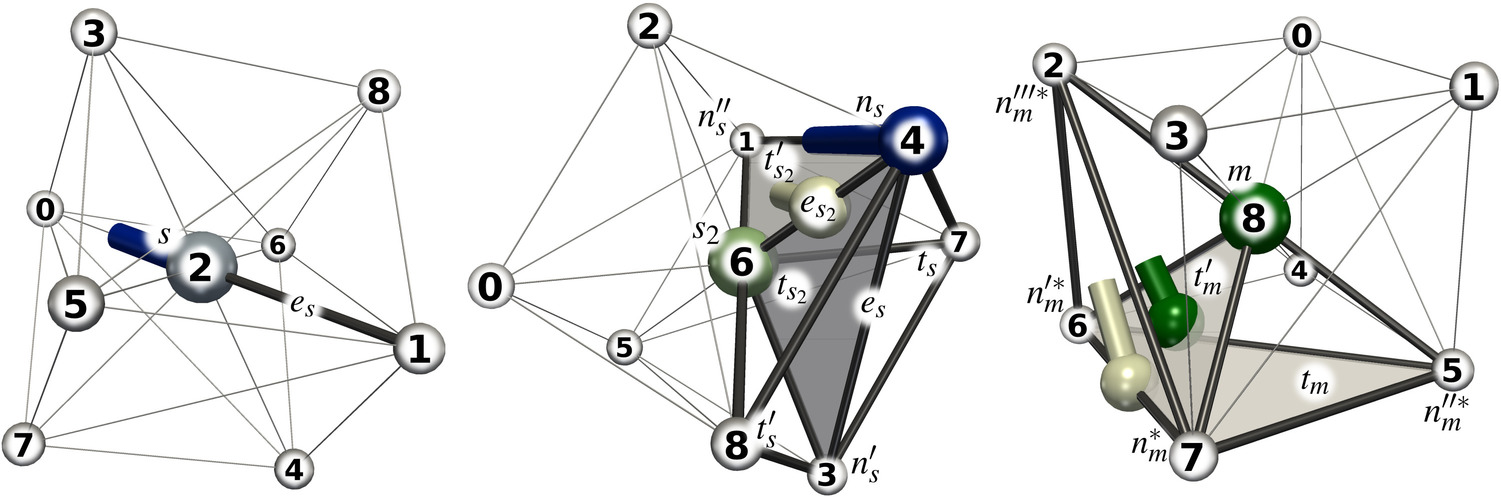}
  \caption{Discrete gradient pairing obtained with 
Alg. 1
in 3D in the star of a PL $1$-saddle 
(left),
  $2$-saddle (center) and maximum (right). Vertex-edge, edge-triangle and 
triangle-tetrahedron pairs are shown with blue, white and green 
balls-and-sticks. Only a few pairs is shown to avoid occlusion.}
  \label{fig_3Dmatching}
\end{figure}

\newpage

\section*{Appendix B: VTK pipeline integration}

A notable software engineering challenge was 
the seamless integration of our
cached triangulation data structure (\texttt{ttk::Triangulation}) into VTK's 
pipeline. 
A naive strategy consists in storing one instance within each VTK wrapper  
(Sec. 
6.2). 
However, this  would duplicate the data structures in the frequent situation 
where multiple TTK modules are lined up within a single VTK pipeline. Instead, 
we implemented a strategy which makes each \texttt{ttk::Triangulation} object 
travel through each pipeline branch without data copy. In particular, for each 
VTK class supported by TTK (\texttt{vtkUnstructuredGrid}, \texttt{vtkPolyData}, 
\texttt{vtkImageData}, etc.), we derived by inheritance a TTK specialization 
(\texttt{ttkUnstructuredGrid}, \texttt{ttkPolyData}, \texttt{ttkImageData}, 
etc.) which holds a pointer to a \texttt{ttk::Triangulation} object. This 
pointer is copied upon VTK's \texttt{ShallowCopy()} operation and a new object 
is actually allocated upon a VTK \texttt{DeepCopy()} operation. On the latter 
operation, primitive-type pointers to the point and simplex lists ($\pointList$ 
and 
$\simplexList$, 
Sec. 5.2)
are passed to the triangulation 
data structure in the case of meshes, and dimensions are passed for that of 
regular grids. Within each VTK wrapper, if the input is a pure VTK 
object (and not a derived TTK object), it 
is first converted into its matching TTK derived class. Then the pointer 
to the \texttt{ttk::Triangulation} 
is extracted and passed to the base code 
functor (Sec. 6.2). Note that this mechanism is automatically handled by TTK 
and is completely hidden to
developers, who only see more general \texttt{vtkDataSet} 
objects passed as arguments of their VTK wrappers. As a consequence, 
\texttt{ttk::Triangulation} objects are allocated only once per pipeline 
branch, and travel by pointers down this branch, without data copy,
possibly 
progressively extending their lists of internal lookup tables upon the 
precondition calls triggered by the successive TTK modules present in the 
pipeline branch.

\end{document}